\documentclass[aoas,preprint]{imsart}

\pdfoutput=1
\RequirePackage{amsthm,amsmath,amsfonts,amssymb}
\RequirePackage{natbib}
\RequirePackage[colorlinks,citecolor=blue,urlcolor=blue]{hyperref}
\RequirePackage{graphicx}
\RequirePackage{longtable, multirow, bm}
\usepackage[final]{pdfpages}
\startlocaldefs

\theoremstyle{remark}

\endlocaldefs

\begin{document}

\begin{frontmatter}
\thispagestyle{empty} 
\title{Estimating the Stillbirth Rate for 195 Countries using\\
a Bayesian Sparse Regression Model \\with Temporal Smoothing\thanksref{T1}}
\runtitle{Global estimation and projection of stillbirth rate}
\thankstext{T1}{The work was supported by the Bill \& Melinda Gates Foundation, UNICEF, and the National Institute of Environmental Health Sciences of the National Institutes of Health under award number T32ES015459.
The content is solely the responsibility of the authors and does not necessarily represent the official views of the funders. Contact: Zhengfan Wang (zhengfanwang@umass.edu) and Leontine Alkema (lalkema@umass.edu).}

\begin{aug}
\author[A]{\fnms{Zhengfan} \snm{Wang}\ead[label=e1]{}}
\author[B]{\fnms{Miranda J.} \snm{Fix}\ead[label=e2]{}}
\author[C]{\fnms{Lucia} \snm{Hug}\ead[label=e3]{}}
\author[C]{\fnms{Anu} \snm{Mishra}\ead[label=e4]{}}
\author[C]{\fnms{Danzhen} \snm{You}\ead[label=e5]{}}
\author[D]{\fnms{Hannah} \snm{Blencowe}\ead[label=e6]{}}
\author[B,E]{\fnms{Jon} \snm{Wakefield}\ead[label=e7]{}}
\and
\author[A]{\fnms{Leontine} \snm{Alkema}\ead[label=e8]{}}
\address[A]{Department of Biostatistics and Epidemiology,
University of Massachusetts Amherst
\printead{e1}}

\address[B]{Department of Biostatistics,
University of Washington
\printead{e2}}

\address[C]{Division of Data, Analytics, Planning and Monitoring, UNICEF
\printead{e3}}

\address[D]{London School of Hygiene and Tropical Medicine, 
University of London
\printead{e4}}

\address[E]{Department of Statistics,
University of Washington
\printead{e4}}


\end{aug}

\begin{abstract}
\noindent Estimation of stillbirth rates globally is complicated because of the paucity of reliable data from countries where most stillbirths occur. We compiled data and developed a Bayesian hierarchical temporal sparse regression model for estimating stillbirth rates for all countries from 2000 to 2019. The model combines covariates with a temporal smoothing process so that estimates are data-driven in country-periods with high-quality data and determined by covariates for country-periods with limited or no data. Horseshoe priors are used to encourage sparseness. The model adjusts observations with alternative stillbirth definitions and accounts for bias in observations that are subject to non-sampling errors. 
In-sample goodness of fit and out-of-sample validation results suggest that the model is reasonably well calibrated. The model is used by the UN Inter-agency Group for Child Mortality Estimation to monitor the stillbirth rate for all countries.
\end{abstract}


\end{frontmatter}
\clearpage
\tableofcontents
\clearpage

\section{Introduction}
The United Nations Inter-agency Group for Child Mortality Estimation (UN IGME) defines a stillbirth as a baby born with no signs of life at 28 weeks or more of gestation, consistent with the International Classification of Diseases (ICD, \cite{icd11}) definition of a ``late gestation fetal death''. Prior estimates highlighted the large global burden of stillbirths with an estimate 2.6 million stillbirths for the year 2015 (\cite{blencowe_national_2016}). Ending preventable stillbirths is one of the core goals of the UN’s Global Strategy for Women’s, Children’s and Adolescents’ Health from 2016 until 2030 (\cite{kuruvilla_global_2016})  and the Every Newborn Action Plan (ENAP, \cite{world_health_organization_every_2014}).  These global initiatives aim to reduce the stillbirth rate (SBR, the number of stillbirths per 1,000 total births) to 12 or fewer stillbirths per 1,000 births in every country by 2030. 

Monitoring of SBRs is challenging because of data paucity in countries where most stillbirths occur. Estimates of SBRs for a country can be derived from administrative data from registration systems (e.g., civil registration and vital statistics (CRVS) and medical birth and death registries). The reliability of SBR estimates from such data sources depends on the accuracy and completeness of reporting and recording of stillbirths and live births. Not all countries maintain an accurate, timely and complete registration system for stillbirths. As a result, stillbirth data from registries can be biased due to underreporting, misclassification, and other data quality issues. Moreover, in many low- and middle-income countries (LMICs), stillbirths are not reported in registration systems at all. For such countries, stillbirth data can be obtained from health management information systems (HMIS), with limitations similar to the ones mentioned for registration systems: HMIS stillbirth data are subject to different stillbirth definitions, can be biased due to underreporting, misclassification and other data quality issues. Lastly, SBR data can be obtained from household surveys and population-based studies but those sources are typically not available for all years of interest and are subject to different stillbirth definitions, potentially large biases and/or non-sampling errors.

\cite{blencowe_national_2016} produced estimates of the SBR for all countries, from 2000 to 2015. Yearly estimates for developed countries with high quality data were obtained from the data directly, using a Loess smoother. Estimates for all other countries were obtained from a regression model with country-specific intercepts and global regression coefficients.  The main limitation of this work is the use of the regression model for countries with limited data: resulting trend estimates are covariate-driven, even if available data suggests deviations away from covariate-predicted trends. In addition, a stepwise approach was taken to carry out variable selection, which underestimates uncertainty since the model selection process is not accounted for.

In this paper, we propose a new approach to estimating the SBR for all countries, using a Bayesian hierarchical temporal sparse regression model (BHTSRM). The model is used by the UN IGME to monitor the stillbirth rate for all countries. Our approach updates and extends the work of \cite{blencowe_national_2016}. As its name implies, BHTSRM combines a hierarchical regression model with a temporal smoothing process. This type of model produces estimates that track high quality data while producing covariate-driven trend estimates for countries with limited or no SBR data. While this kind of model has been used for estimating global health indicators in other settings, e.g. in \cite{alkema_bayesian_2017}, prior work does not address variable selection in this context. 
Here we extend upon prior work by introducing sparsity-inducing priors for estimating regression coefficients. In particular, we use horseshoe priors (\cite{piironen_sparsity_2017}) to shrink the less important coefficients toward zero. 

Our proposed model also introduces new statistical approaches to address various data quality issues. Firstly, we propose a procedure for data exclusion based on comparing observed ratios of SBR to the neonatal mortality rate (NMR) for the population of interest, and excluding ratios that suggest that stillbirths are underreported as compared to neonatal deaths. Secondly, we introduce an estimation approach to incorporate observations with alternative definitions of a stillbirth (e.g., based on 22 weeks gestational age or 1000 grams birthweight) into the model while accounting for additional uncertainty associated with such observations. 

This paper is organized as follows: in Section \ref{sec-data-availability}, we provide an overview of data sources and definitions that are available for measuring SBR and Section~\ref{sec-ratio} introduces the exclusion of data based on the ratio of SBR to NMR. We describe the BHTSRM in Section \ref{sec_methods}. In Section \ref{sec-result}, we present estimates of SBR, data quality parameters and validation results. Last, we conclude with a discussion of limitations and future research directions in Section \ref{sec-discussion}.

\section{Data}\label{sec-data}
\subsection{Data availability}\label{sec-data-availability}
Estimates of SBRs for a country were compiled from various sources, including administrative data (e.g. CRVS and birth or death registries), HMIS, household surveys, and population-based studies obtained from a review of the academic literature. Data from all available sources were compiled by the UN IGME. Data processing and general exclusion steps are described in detail in the Appendix Section~\ref{sec-dataexclusion}. Exclusion based on the ratio of SBR to NMR is described in Section~\ref{sec-ratio}. After exclusion, we used 1531 observations to produce estimates for 195 countries.

\begin{table}[]
\begin{tabular}{l|cc|cc}
\hline
  \multirow{2}{*}{Data Source}      & \multicolumn{2}{c|}{Data Available} & \multicolumn{2}{c}{Data Used in Model} \\ \cline{2-5} 
                   & Number of Countries       & Number of Obs       & Number of Countries         & Number of Obs        \\ \hline
Administrative              & 105                 & 1738          & 75                    & 1157          \\
HMIS               & 53                  & 506           & 26                    & 162            \\
Population Study & 35                  & 363           & 23                    & 117            \\
Household Survey   & 73                  & 226           & 44                    & 95             \\ \hline
Total              & 171                 & 2833          & 133                   & 1531           \\ \hline
\end{tabular}
\end{table}

\begin{table}[]
\begin{tabular}{l|cc|cc}
\hline
  \multirow{2}{*}{Definition}      & \multicolumn{2}{c|}{Data Available} & \multicolumn{2}{c}{Data Used in Model} \\ \cline{2-5} 
                   & Number of Countries       & Number of Obs       & Number of Countries         & Number of Obs        \\ \hline
28 weeks              & 154                & 2067          & 124                    & 1220          \\
24 weeks               & 8                  & 49           & 3                    & 44            \\
22 weeks     & 25                  & 219           & 15                    & 85            \\
1000 grams   & 21                  & 166           & 20                    & 146             \\ 
500 grams   & 8                  & 65           & 5                    & 36             \\ 
other   & 38                  & 267           & N/A                    & N/A             \\ \hline
Total              & 171                 & 2833          & 133                   & 1531           \\ \hline
\end{tabular}
\label{tb-defdata}
\caption{Data availability by data source and definition for countries in 2000-2019. Data in one country might be available in multiple definitions and come from multiple data sources.}\label{tb:datoverview}
\end{table}

To allow for international comparison, we focus on estimating SBRs using the observations based on the $\geq 28$ weeks definition. However, not all data sources provide information based on this definition, for example stillbirths may be reported by birthweight or an alternative gestational age cut-off, see Table~\ref{tb:datoverview}. In fitting the SBR model, we used data based on the $\geq 28$ weeks definition when available. Otherwise, observations recorded using alternative definitions were adjusted to the 28 weeks definition which introduces additional uncertainty associated with the alternative definition (see Section~\ref{sec-def}).

Data availability (after data exclusion) is illustrated for selected countries in Figure~\ref{fig:ill2}. Data availability ranges in the selected countries from no included data in Afghanistan to an annual time series of national administrative data based on the 28 week definition for Ireland. Botswana, Malawi, Uganda and Ukraine are examples of countries with SBR data from multiple sources, available for selected periods only. In Ukraine, SBR data are available from 2007 to 2017 from administrative systems but recorded using a 22 week definition. In Uganda, the only available data comes from surveys and population-based studies. In Malawi, available data sources are HMIS, population-based studies, and household surveys.

\clearpage 

\begin{figure}[htbp]
    \centering
       \includegraphics[page=1,width=0.85\textwidth,height=0.32\textheight]{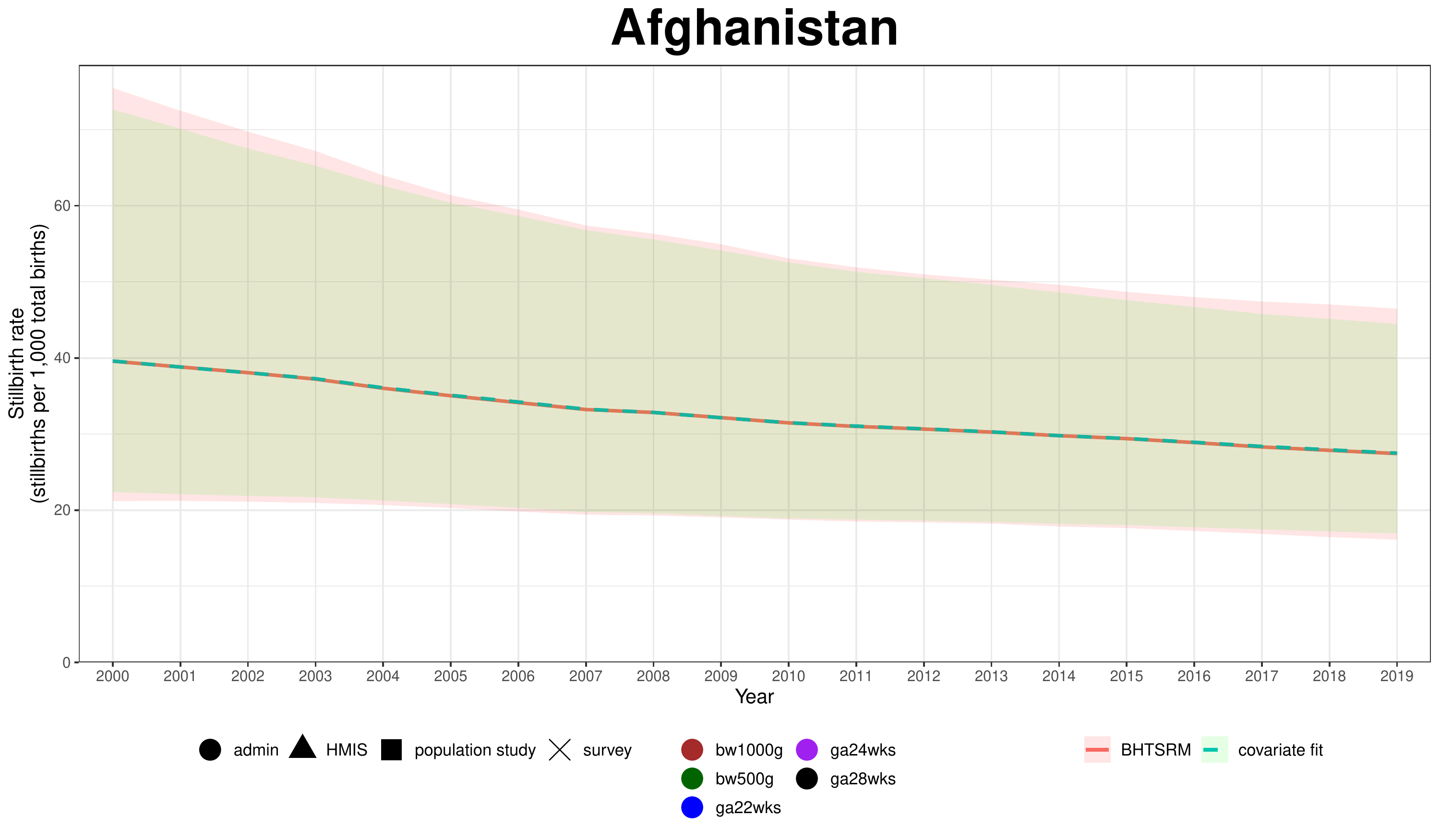}\\
         \includegraphics[page=28,width=0.85\textwidth,height=0.32\textheight]{fig/compar_plot_final.pdf}\\
                  \includegraphics[page=80,width=0.85\textwidth,height=0.32\textheight]{fig/compar_plot_final.pdf}\\

-- Figure continued on next page --
\end{figure}
\clearpage 
\begin{figure}[htbp]
    \centering
               \includegraphics[page=123,width=0.85\textwidth,height=0.32\textheight]{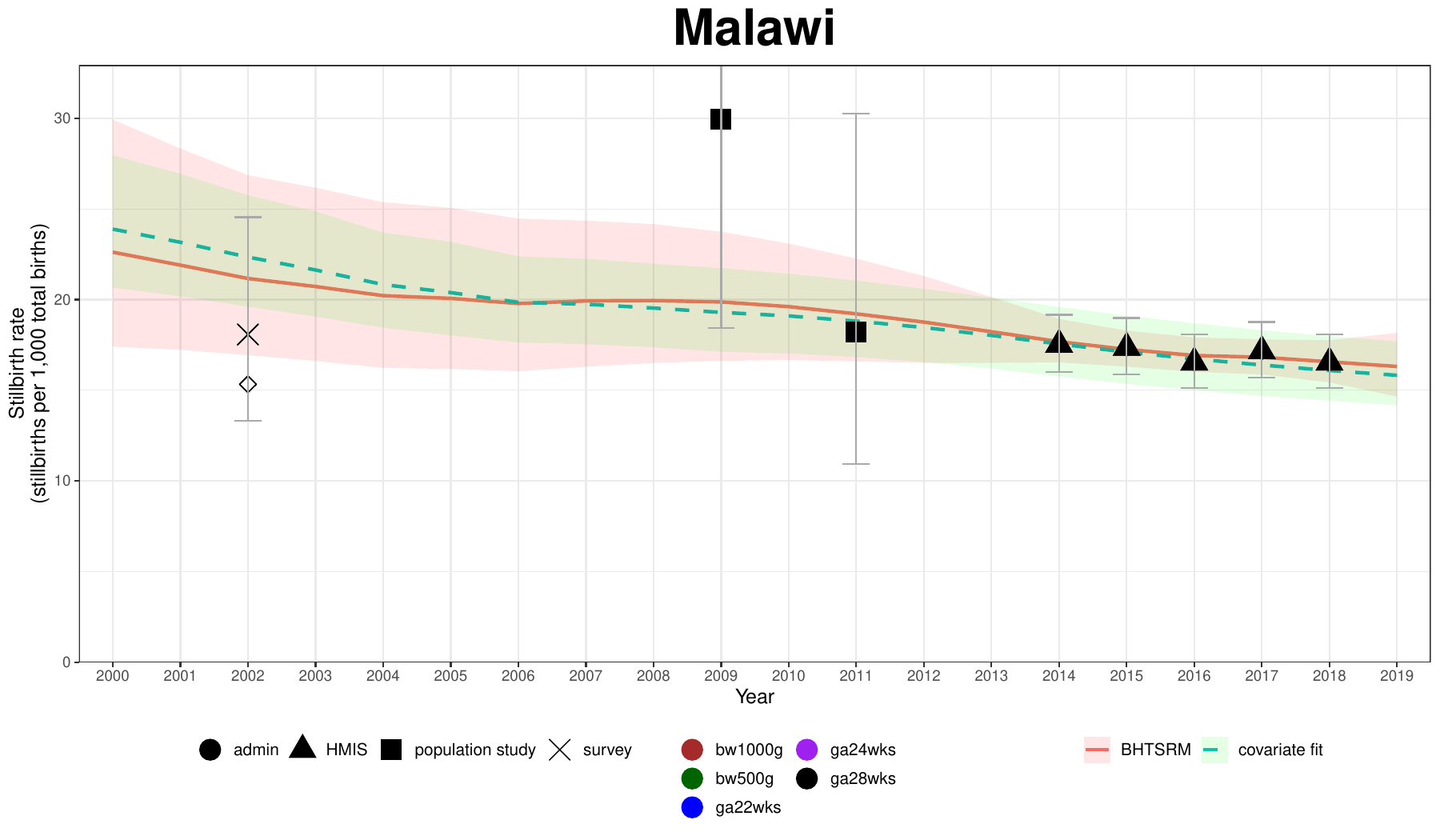}\\
       \includegraphics[page=182,width=0.85\textwidth,height=0.32\textheight]{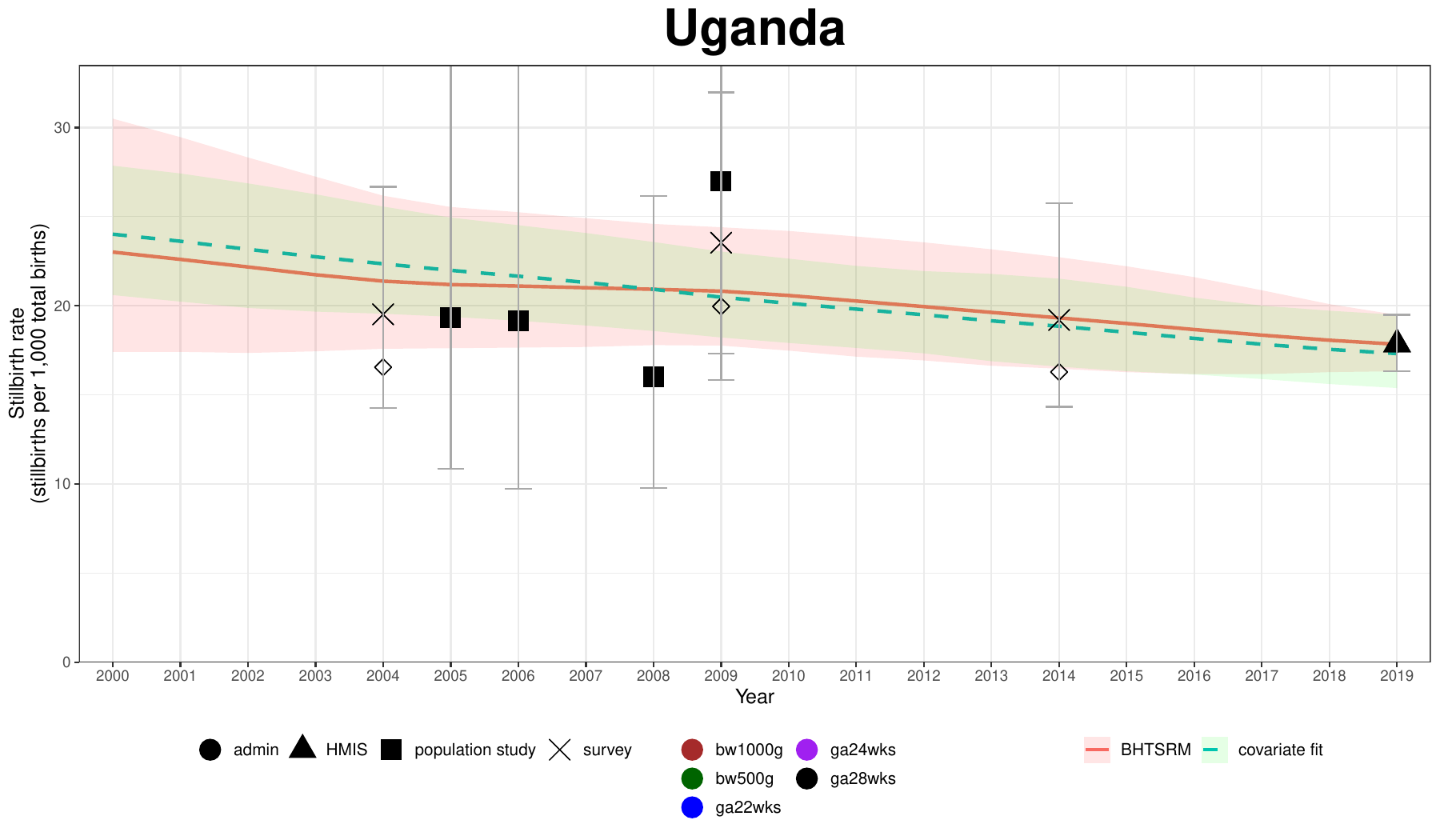}\\
           \includegraphics[page=183,width=0.85\textwidth,height=0.32\textheight]{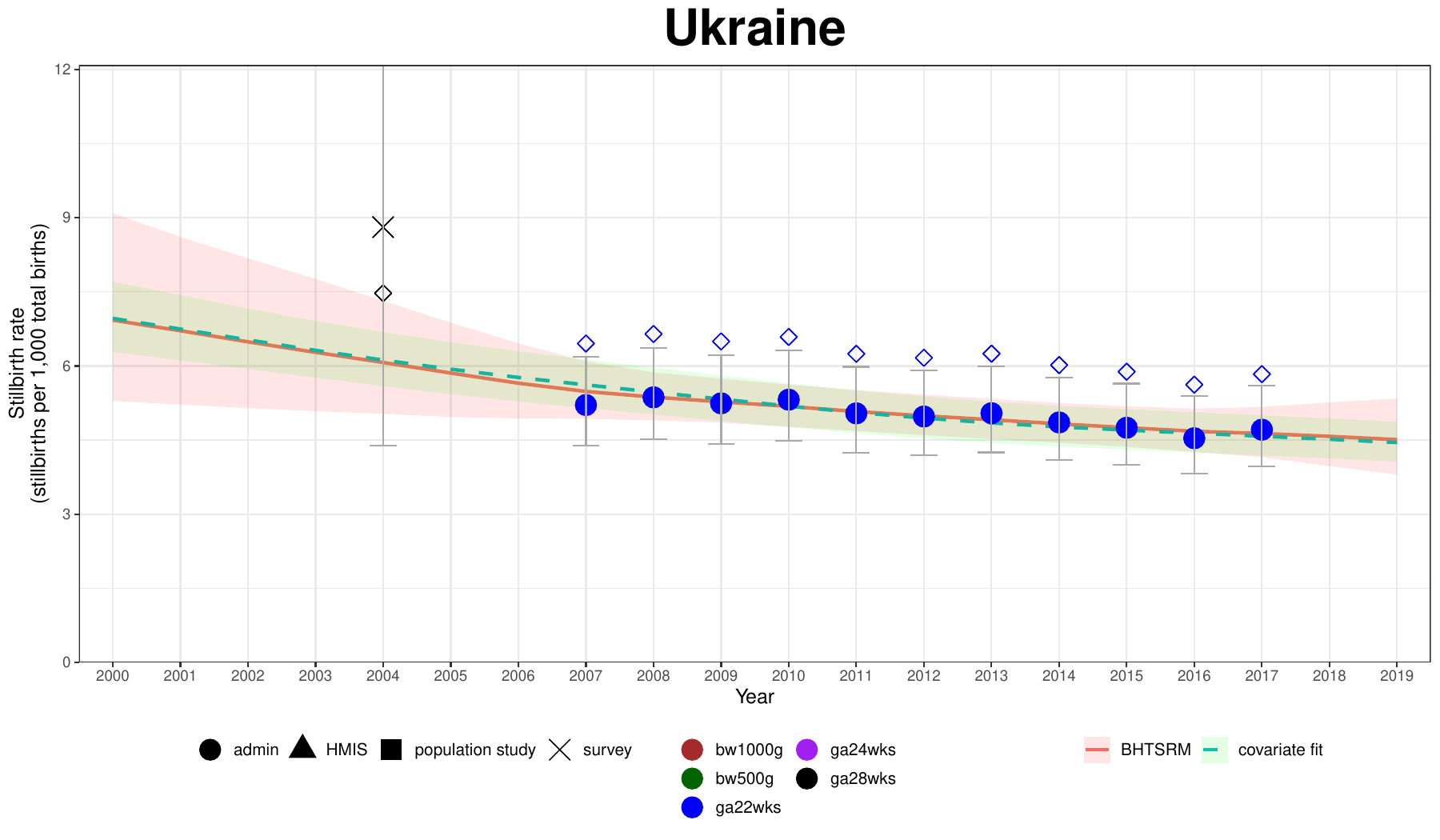}\\
           -- Figure caption on next page --
\end{figure}

\clearpage 
\begin{figure}[htbp]
    \caption{SBR data and estimates for 2000-2019 for selected countries. Posterior median point estimates from BHTSRM (red line) and 90\% credible intervals (red area), and covariate-based estimate (dashed green line) and 90\% credible intervals (green area) are shown. Observed but unadjusted observations are displayed by hollow symbols. Adjusted data (based on definitional adjustments and accounting for survey biases where applicable) are shown for all source types. Colors indicate the definition of the observation. Error bars displayed with adjusted observations indicate 95\% confidence interval of the SBR based on its estimated bias and error variance. Note that the y-axis varies across countries, and that data excluded based on the data quality assessment are not shown. }
    \label{fig:ill2}
\end{figure}

\subsection{Exclusion based on the ratio of SBR to NMR}\label{sec-ratio}
Stillbirths may be underreported for various reasons, including lack of understanding of the definition of a stillbirth, lack of motivation for reporting these events amongst health workers, data collectors and bereaved parents, fear of litigation or disciplinary action, as well as distress, cost, stigma or other negative factors associated with reporting (\cite{christou_how_2019}). Generally, stillbirths are more poorly recorded than deaths of liveborn neonates, which are themselves under-recorded in many settings (\cite{stanton_stillbirth_2006} and \cite{woods_long-term_2008}).

Similar to the work of \cite{blencowe_national_2016}, we excluded datapoints that are likely to reflect poor case ascertainment based on an assessment of the ratio of SBR to NMR. We describe the approach in detail in the remainder of this section. In summary, we assume that each observed log-ratio is the sum of a setting-specific expected log-ratio and random error. We use high quality data from LMIC to estimate the mean and variance of the distribution of expected log-ratios. Intuitively, if case ascertainment is poor, the ratio of SBR to NMR would be small. We calculate observed log-ratios for all observations in the data set and exclude observations that are deemed subject to underreporting. Specifically, we build a model for the log-ratio and if the observed ratio lies in the lower (5\%) tail of its predictive distribution, we deem the observation subject to underreporting and exclude the datapoint. 

\subsubsection{Analysis of SBR to NMR ratios in high quality data}
High quality data collected in LMICs is used to analyze the distribution of the ratio of SBR (measured as per 28 week definition) to NMR. The high-quality data are population-based prospectively-collected data with recruitment prior to 28 weeks of gestation, and follow-up to at least 28 days of live births (\cite{bose_global_2015} and \cite{ahmed_population-based_2018}). Let $r_i = y_i/o_i$ denote the observed ratio of SBR $y_i$ to NMR $o_i$. We assume that each observed log-ratio is the sum of a setting-specific expected log-ratio and random error:
\begin{eqnarray}
\log(r_i) &=& \theta_i + \varepsilon_i, \label{eq-ratio}\\
\varepsilon_i &\sim& N(0, v_i^2), \label{eq-ratio2}
\end{eqnarray}
where $\theta_i = E(\log r_i)$ refers to the expected log-ratio of SBR to NMR, and $\varepsilon_i$ refers to random error with mean zero and variance $v_i^2$ which is assumed known (see Appendix section~\ref{sec-app-ex} for a full description of how this variance term is approximated). We assume for the expected log-ratios $\theta_i$:
\begin{eqnarray}
\theta_i|\mu_{\theta}, \sigma_{\theta}^2  &\sim& N(\mu_{\theta}, \sigma_{\theta}^2),\label{eq-ratio3}
\end{eqnarray}
where $\mu_{\theta}$ represents the mean log-ratio across different SBR and NMR settings and $\sigma_{\theta}^2$ refers to variability across settings. We assign vague priors to $\mu_{\theta}$ and $\sigma_{\theta}^2$.

\subsubsection{Exclusion procedure}  
We calculate observed SBR to NMR ratios for all observations in the data set, using SBR and NMR from the same data source. Where data sources have missing NMR data, national NMR estimates from the UN IGME  are used. For observations from HMIS and population-based studies on stillbirths, the ratio of observed SBR to the UN IGME NMR is calculated and the same exclusion approach applied so that observations with extremely low SBR compared to national level NMR are excluded. 

If stillbirths are underreported relative to neonatal deaths for a specific observation, its associated observed log-ratio of SBR to NMR $r_i$ is biased downwards as compared to the true log-ratio $\theta_i$. We exclude observations that are deemed subject to underreporting based on the 1-sided tail probability of observing a value more extreme than the reported ratio $r_i$. Specifically, we calculate
$$p_i = \int^{\log(r_i)}_{-\infty} f( r) dr,$$
where $f( r)$ is the predictive density for $\log(r_i)$ based on Equation~(\ref{eq-ratio}), using observation-specific error variance and point estimates for the mean and across-setting variance of $\theta_i$. We  exclude observation $i$ if $p_i < 0.05$, i.e. if the probability of observing a ratio more extreme than $r_i$ is less than 5\%. For the data with alternative stillbirth definition (non 28 weeks definition), we apply the exclusion procedure after definition adjustment (see Section~\ref{sec-def}).

\section{Methods}\label{sec_methods}

\paragraph*{Notation}
We use lowercase Greek letters for unknown parameters and uppercase Greek letters for variables which are functions of unknown parameters (modeled estimates). Roman letters indicate variables that are known or fixed, including data (in lowercase) and estimates provided by other sources or the literature (in uppercase).

We let $\Omega_{c,t}$ denote the main outcome of interest, which is the SBR for country $c$ in year $t$ for $\ge 28$ weeks gestational age. Observations are available across countries over time and are indexed by $i\in\{1, \cdots, n\}$; $c[i]$ refers to the country for which the $i$-th observation was recorded, $t[i]$ the calendar year of the observation, $j[i]$ the data source type of the observation, and $d[i]$ to its stillbirth definition. The index $r[c]$ refers to the region of country $c$.

\subsection{SBR model summary}
Let $y_i$ denote an observed SBR for country $c[i]$, in year $t[i]$. We assume the following data model: 
\begin{equation}
\log(y_i)|\Theta_{c[i],t[i]}, \psi_{j[i]}, \sigma_{j[i]}^2 \sim N(\Theta_{c[i],t[i]} + \psi_{j[i]} + \hat{\gamma}_{d[i]}, s_i^2 + \hat{\varphi}_{d[i]}^2 + \sigma_{j[i]}^2),\label{eq-dm}
\end{equation}
where  $\Theta_{c,t} = \log(\Omega_{c,t})$ refers to the log-transformed true SBR $\Omega_{c,t}$ for that country-year,  $s_i^2$ to variance of $\log(y_i)$, $\psi_{j[i]}$ and $\sigma_{j[i]}^2$ refer to its source type-specific bias and variance, respectively, 
and $\hat{\gamma}_{d}$ and $\hat{\varphi}_{d}^2$ to definition-specific adjustment and variance for observations that are reported using definitions other than 28 weeks of gestational age. Definitional adjustment parameters are estimated prior to model fitting, as discussed in Section~\ref{sec-def}. The process model specification for $\Theta_{c,t}$ is explained in Section~\ref{sec-process}. 

The term $s_i^2$ refers to the variance of $\log(y_i)$. Calculation of these variances is outlined in Section~\ref{sec-se}. Source type bias terms $\psi_{j}$ are included in model fitting to capture systematic biases associated with specific source types. We assume there is no source type biases for administrative, HMIS, and population-based studies, i.e.
$\psi_j = 0$ for $j$ referring to these three source types. \cite{liu_understanding_2016} and \cite{Bradley_contraceptive_2015} suggest that stillbirths tend to be underreported in surveys, so we assume that data from surveys have a negative bias term and estimate this bias term. 
The measurement error variance term $\sigma_j^2$ captures non-systematic errors due to errors introduced in reporting. These variance parameters are estimated and assigned vague priors (see Appendix Section~\ref{sec-app-eqs}).

\subsection{Definitional adjustment}\label{sec-def}

To allow for international comparison, we focus on estimating SBRs using the definition of stillbirth as any baby born without signs of life at greater than or equal to 28 weeks of completed gestation. However, not all data sources provide information based on this definition, for example stillbirths may be reported by birthweight or an alternative gestational age cut-off. In fitting the model, we use data based on the $\geq 28$ weeks definition when available. Otherwise, observations recorded using alternative definitions are used by adjusting for the different definitions and additional uncertainty associated with the alternative definitions.

We use data sources that reported stillbirth using multiple definitions to estimate the relationships between SBRs based on 28 week gestational age and alternative definitions. We assume that the true log-transformed SBR for observation $i$ under definition $d[i]$, $ \Theta_{c[i],t[i]}^{(d)}$, is
\begin{align}\label{}
\Theta_{c[i],t[i]}^{(d)} &= \Theta_{c[i],t[i]} + \kappa_i^{(d)}, \label{eq:kappa}
\end{align}
where $\Theta_{c[i],t[i]}$ refers to the log SBR under the 28 week definition. For adjustment in the SBR model, we set adjustment ${\gamma}_{d}$ and variance ${\varphi}_{d}^2$ equal to the posterior median and variance of the predictive distribution for $\kappa_i^{(d)}$ for each alternative definition $d$.

To estimate the definition-specific adjustment and variance (i.e., to obtain $\hat{\gamma}_d$ and $\hat{\varphi}_d^2$), we use data sources that reported stillbirths using multiple definitions. For a given alternative definition $d$, we look at paired observations of stillbirth counts $(z_i^{(d)}, z_i)$, where $z_i$ is the number of stillbirths under the $\geq 28$ weeks definition, $z_i^{(d)}$ is the number of stillbirths under the alternative definition $d$, and $i = 1, \dots, n_d$, where $n_d$ is the number of pairs available for definition $d$ from the same source, country, and year, under that definition and $\geq 28$ weeks definition. We use the paired counts to estimate ${\gamma}_{d}$ and ${\varphi}_{d}^2$ for definition $d$ separately for high-income countries (HICs) and LMICs.

Data limitations necessitate some assumptions regarding definitional adjustments. For survey data, a seven-month duration of pregnancy is assumed to be equal to a 28 weeks or more definition. In HICs, we are able to use paired observations to estimate the adjustment and variance associated with the following definitions: $\geq 22$ weeks, $\geq 24$ weeks, $\geq 500$ g, and $\geq 1000$ g.
Using high-quality study data in LMICs, we estimate the adjustment and variance associated with the $\geq 22$ weeks definition. Due to lack of other data, in LMICs we assume that $\geq 500$ grams birthweight is equivalent to $\geq 22$ weeks of gestational age, and $\geq 1000$ grams birthweight is equivalent to $\geq 28$ weeks of gestational age.

Alternative definitions fall into two categories: definitions containing the $\geq 28$ weeks definition and definitions overlapping with the $\geq 28$ weeks definition. We consider each of these below. Additional details (including choice of priors) are given in Appendix Section~\ref{sec-app-def}. 

\subsubsection*{Definitions containing the $\geq 28$ weeks definition}
Stillbirths $z_i$ recorded using the $\geq 28$ weeks definition are a subset of stillbirths recorded using the $\geq 22$ or $\geq 24$ weeks definitions, $z_i \leq z_i^{(d)}$ for $d$ referring to 22 and 24 weeks. We assume
\begin{align}
z_i | \omega_i^{(d)} &\sim \text{Binomial}(z_i^{(d)}, \omega_i^{(d)}), \label{eq:subsetBinom}\\
\text{logit}(\omega_i^{(d)})|\mu_{\omega, d}, \sigma_{\omega, d}^2 &\sim N(\mu_{\omega, d}, \sigma_{\omega, d}^2),
\end{align}
where $\omega^{(d)}$ is the definition-specific probability of a stillbirth with gestational age of $\geq$ 22 or $\geq$ 24 weeks being born dead after 28 weeks. For using these observations in SBR model fitting, we use the approximation $\kappa_i^{(d)} \approx -\log(\omega_i^{(d)})$, based on approximate equality between the ratio of stillbirths and the ratio of SBRs (see Appendix Section~\ref{sec-app-def}). Estimates for the adjustment $\hat{\gamma}_d$ and variance $\hat{\varphi}_d^2$ in Equation~\eqref{eq-dm} are given by the median and variance of  the predictive distribution for $-\log(\omega)_i^{(d)}$. 

\subsubsection*{Definitions overlapping with the $\geq 28$ weeks definition}
Stillbirths $z_i^{(d)}$ recorded using the $\geq 1000$ or $\geq 500$ grams definitions are overlapping with the stillbirths $z_i$ using the $\geq 28$ weeks definition.
In this setting, let $z_i^{(d)} = a_i + b_i$ and $z_i = a_i + c_i$, where $a_i$ is the count of stillbirths that satisfy the 28-week and alternative definition, $b_i$ is the counts of stillbirth with alternative definition rather than 28 weeks definition, $c_i$ is the count of stillbirth with 28 weeks definition rather than alternative definition, so that $N_i = a_i + b_i + c_i$ is the total number of stillbirth based on the 28 weeks definition or an alternative definition. We assume 
\begin{align}
    (a_i,b_i,c_i)|(\omega_{i,a}^{(d)},\omega_{i,b}^{(d)},\omega_{i,c}^{(d)}) \sim Multinom\left(N_i,(\omega_{i,a}^{(d)},\omega_{i,b}^{(d)},\omega_{i,c}^{(d)})\right),
\end{align}
where $\omega_{i,a}^{(d)}, \omega_{i,b}^{(d)}$ and $\omega_{i,c}^{(d)}$ refer to the probability of a stillbirth of the 28-week or alternative definition satisfying both definitions, the 28-week only, or the alternative definition only. For using these observations in SBR model fitting, we use the approximation  $ \kappa_i^{(d)} \approx \Gamma_i^{(d)}$ (see Appendix Section~\ref{sec-app-def}) with  $\Gamma_i^{(d)}$ referring to the  log-ratio of the 28 weeks to alternative definition $d$:
\begin{align}
    \Gamma_i^{(d)}  = \log\left(\frac{\omega_{i,a}^{(d)}+\omega_{i,b}^{(d)}}{\omega_{i,a}^{(d)}+\omega_{i,c}^{(d)}}\right).
\end{align}
We assume 
\begin{align}
    \Gamma_i^{(d)}|\mu_{\Gamma,d },\sigma_{\Gamma, d}^2 \sim N(\mu_{\Gamma,d },\sigma_{\Gamma, d}^2).
\end{align}
We obtain estimates for the adjustment $\gamma_d$ and variance $\varphi_d^2$ in Equation~\eqref{eq-dm} using the median and variance of the predictive distribution for $\Gamma_i^{(d)}$.

\subsection{Bayesian hierarchical temporal sparse regression model}\label{sec-process}

We developed a Bayesian hierarchical temporal regression model (BHTRM) to estimate the SBR for all country-years. BHTRMs have been developed for the estimation of demographic indicators for multiple countries and years, for example for maternal mortality (\cite{alkema_bayesian_2017}). The BHTRM combines a regression model with a temporal smoothing process. Specifically, a BHTRM for $\Theta_{c,t}$, the true log-transformed  SBR ($\ge 28$ weeks gestation) for country $c$ and year $t$, is defined as follows:
\begin{equation}
\Theta_{c,t} = \varsigma_c  + \sum_k X_{k,c,t} \beta_k + \delta_{c,t},
\end{equation}
were $\varsigma_c$ refers to the country-specific intercept, $\sum_k X_{k,c,t} \beta_k$ refers to the linear regression function and $\delta_{c,t}$ refers to a temporal smoothing process. Using BHTRMs, estimates can track high quality data and fall back to levels and rates of change implied by the covariates if there are no data or if data are too uncertain. In limited data setting, estimates will be uncertain due to uncertainty in the smoothing component $\delta_{c,t}$. The remainder of this section presents the model set. Information on prior distributions is given in Section~\ref{sec-app-eqs}.  

\paragraph*{Country-specific intercepts}
 The country-specific intercept $\varsigma_c$ is estimated hierarchically, with 
\begin{eqnarray}
\varsigma_c|\eta_{r[c]},\sigma_{\varsigma}^2 &\sim& N(\eta_{r[c]},\sigma_{\varsigma}^2),\\
\eta_r|\xi_w,\sigma_{\eta}^2 &\sim& N(\xi_w,\sigma_{\eta}^2),
\end{eqnarray}
where $r[c]$ refers to the region of country $c$ (based on 6 regions, $r=1,...,6$), $\eta_{r}$  refers to the regional mean,  $\sigma_{\varsigma}^2$ to the across-country variance within regions,  $\xi_w$ to the global mean, and $\sigma_{\eta}^2$ to the across-region variance. Vague priors are used for the variances and global mean.

\paragraph*{Temporal smoothing process}
For SBR estimation, a penalized spline regression model is used for $\delta_{c,t}$, defined as follows:
\begin{equation}
\delta_{c,t} =  \sum_{h=1}^H k_h(t)\alpha_{h,c},
\end{equation}
where $k_h(t)$ refers to the $h$-th spline function evaluated at time $t$ and $\alpha_{h,c}$ to its regression coefficient for country $c$. We use equally spaced quadratic B-splines, with knots spaced 1 year apart and placed at each integer year (\cite{eilers_flexible_1996} , \cite{currie_flexible_2002}). 
The spline regression coefficients are modeled with a first-order random walk process with a sum-to-zero constraint $\frac{1}{H} \sum_h \alpha_{h,c} = 0$ to ensure identifiability. For each country, we define first order difference $\Delta \alpha_{h,c}$:
\begin{eqnarray}
\Delta \alpha_{h,c} &=& \alpha_{h,c} - \alpha_{h-1,c}. 
\end{eqnarray}
First-order differences are penalized as follows
\begin{eqnarray}
\Delta \alpha_{h,c}|\sigma_{\Delta}^2 \sim N(0,\sigma_{\Delta}^2),
\end{eqnarray}
where the variance term $\sigma_{\Delta}^2$ determines the smoothness of the fit.

\paragraph*{Estimating regression coefficients using sparsity-inducing priors}
\cite{blencowe_national_2016} identified a large number of candidate covariates for estimating SBR based on a conceptual framework. The framework includes distal determinants such as socio-economic factors, demographic and biomedical factors, associated perinatal outcome markers, and access to health care. In the prior SBR estimation study, a stepwise approach was used for variable selection (\cite{blencowe_national_2016}). In this study, we refrain from stepwise selection methods and instead use regularized horseshoe priors on regression coefficients (\cite{piironen_comparison_2017}) to impose sparsity, i.e., by shrinking a subset of coefficients toward zero.  We expand upon BHTRMs by introducing sparsity-inducing priors for estimating regression coefficients $\beta_k$ and refer to the resulting model set up as a Bayesian hierarchical temporal sparse regression model (BHTSRM).

Regularized horseshoe priors for the regression coefficients are defined as follows :
\begin{eqnarray}
    \beta_k|\lambda_k,\tau,\rho &\sim& N(0,\tau^2\tilde{\lambda}_k^2),\\
    \tilde{\lambda}_k^2 &=& \frac{\rho^2\lambda_k^2}{\rho^2+\tau^2\lambda_k^2},\\
    \lambda_k &\sim&  C^+(0,1),
    \end{eqnarray}
where $\tau$ and $\rho$ are global shrinkage parameters, and the $\lambda_k$s are local (coefficient-specific) parameters. In this set-up, the global hyperparameter $\tau$ shrinks all the parameters towards zero, while the heavy-tailed half-Cauchy prior $C^{+}(0,1)$ for the coefficient-specific parameters $\lambda_k$s allow some $\beta_k$s to escape from the global shrinkage. 
We use the following priors for $\tau$ and $\rho$ (\cite{piironen_sparsity_2017}):
\begin{eqnarray}
\tau &\sim& C^+ (0,\tau_0),\\
    \rho^2 &\sim& \text{Inv-Gamma}(q,g),
\end{eqnarray}
where $\tau_0$, $q$ and $g$ are fixed. 
We set $\tau_0=1$, $q = 2$, and $g=8$, as per the recommended defaults in  \cite{piironen_sparsity_2017}, \cite{carvalho_handling_nodate}, and \cite{gelman_prior_2006}. We address the sensitivity to these settings in Appendix Section~\ref{sec-modelcompare}. 

\paragraph*{Subsetted model for producing UN IGME SBR estimates}\label{sec-subset}
While the BHTSRM results in a subset of regression coefficients that are shrunk toward zero, it does not result in a sparse model in the sense of having regression coefficients that are equal to zero. Conditional on comparable model performance, a more parsimonious model is preferable over the BHTSRM for the UN IGME to produce country estimates for simplicity of use and communication. We obtain a subsetted model from the BHTSRM model fit for model comparison and, after finding that the subsetted model and the BHTSRM produce similar results, we use the subsetted model for producing SBR estimates. 

The covariates included in the subsetted model are obtained based on the absolute size of the regression coefficients. After fitting the BHTSRM, we order the covariates by their absolute regression coefficient and apply an absolute cut off to construct the simpler subset. We then refit the subsetted model to the data, using vague priors for the regression coefficients:
\begin{eqnarray}
    \beta_k &\sim& N(0,5^2).
\end{eqnarray}

We compare the two model fits (all covariates with horseshoe priors and subsetted model) in terms of differences in country estimates and approximate leave-one-out-validation.

\subsection{Computation}
The Hamiltonian Monte Carlo (HMC) algorithm is employed to sample from the posterior distribution of the parameters with the use of Stan (\cite{carpenter_stan_2017}) and R package Rstan (\cite{rstan_package}). Six parallel chains are run with a total of 6,000 iterations in each chain. The first 2000 iterations in each chain are discarded as burn-in so that the resulting chains contain 4,000 samples each. 
Standard diagnostic checks are used to check convergence and sampling efficiency. These checks are based on trace plots, the improved Gelman and Rubin diagnostic (\cite{mcmc_diag_GR}, 
\cite{vehtari_rank-normalization_2020}), and various calculations of effective sample size (ESS), including the bulk ESS - using rank-normalized draws - and the tail ESS - giving the minimum of the effective sample sizes of the 5\% and 95\% quantiles (see \cite{STANwarnings} for further details).

\subsection{Model validation}

Model performance is assessed through two out-of-sample validation exercises. In the first exercise, we randomly leave out 20\% of the observations and repeat this exercise 20 times (leaving out 306 observations each time). In the second exercise, we leave out the last observation for each country to check the predictive performance. To evaluate model performance, we calculate various measures based on a comparison between left-out observations and their predictive distributions. We define prediction errors $e_i$ as the difference between the left-out observation and the median of its predictive posterior distribution based on the training set: 
\begin{equation}
    e_i = (\log(y_i) - \log(\tilde{y_i}))/S_i,
\end{equation}
where $y_i$ is the left-out observations, and $\tilde{y_i}$ and $S_i$ refer to the estimated median and standard deviation of the predictive distribution for $y_i$ based on the training set. Coverage of prediction intervals is given by $N^{-1} \sum_{i=1}^{N} 1[l_i \leq y_i \leq  u_i]$, where $N$ denotes the total number of left-out observations considered, and $l_i$ and $u_i$ are the lower and upper bounds of the prediction interval for the $i$th observation. We also carry out approximate leave-one-out cross-validation (LOO), which is implemented in the loo package in R (\cite{vehtari_practical_2017}, \cite{loo_package}).

\section{Results}\label{sec-result}

\subsection{Model validation}

Validation results are given in Table~\ref{tb-val}. For all scenarios, mean residuals are close to zero, suggesting small bias. Mean absolute residuals are around 0.1. The approximate leave-one-out validation exercise suggests that predictive distributions are overdispersed as compared to the left-out observations, with the percentages outside of 80\% and 90\% prediction intervals being lower than expected. The out-of-sample exercises suggest that the model is reasonably well calibrated, with slightly less left-out observations falling below their respective predictive intervals than expected. 

\begin{table}[h]
\centering
\begin{tabular}{|c|c|c|c|c|c|c|c|}
\hline
Validation & Mean error & Mean abs. error  & below 5\% & below 10\% & above 90\% & above 95\% & N.test \\ 
Desirable$^*$ & 0 & N/A & 5\% & 10\% & 10\% &5\%  & - \\ \hline 
Recent       & 0.003       & 0.104  & 3.6\%     & 5.4\%      & 7.1\%      & 4.5\%                 & 112    \\ \hline
Random    & 0.002     & 0.120 &  3.3\%     &  6.4\%      &  10.0\%       &  5.6\%                  & 306    \\ \hline
In-sample    & -0.005        & 0.091  & 1.6\%    & 3.4\%      & 4.3\%      & 1.8\%                 & 1531   \\ \hline 
\end{tabular}
\caption{Validation results for SBR estimates. Validation exercises Recent, Random, and In-sample represent leaving out recent observations, randomly leaving out 20\% of all observations, and approximate leave-one-out validation, respectively. The outcome measures are as follows: mean of error, mean absolute error, and \% of left-out observations below and above their respective 90\% and 80\% prediction intervals. Desirable$^*$ refers to outcomes for models that are unbiased and well calibrated.}
\label{tb-val}
\end{table}

\subsection{Data exclusion based on the SBR to NMR ratio}

We use the LMIC high quality data to analyze the distribution of the ratio of SBR to NMR. The estimated mean ratio on the log scale is $\hat{\mu_\theta} = -0.180$ (-0.250, -0.111) and variance across settings is estimated as $\hat{\sigma}_\theta^2=0.083$. An observation is excluded if its observed ratio is less than the 5\% lower bound of its corresponding predictive distribution of the SBR to NMR ratio as described in Equations~(\ref{eq-ratio}) - (\ref{eq-ratio3}). Based on the posterior samples of ${\mu_\theta}$ and ${\sigma}_\theta^2$, the 5\% lower bound of the predictive distribution of the SBR to NMR ratio is 0.52 for observations with variance $v_i=0$.  We apply this exclusion approach to all observations in the raw input data base with 28 weeks definition, and to the adjusted observations with alternative definitions.

\subsection{Data quality and data adjustments}
Adjustments and error variances associated with different definitions are given in Table \ref{tb-def}. Adjustments on the log-scale for high income countries range from -0.065 (-0.074, -0.056) for definitions in the 1000 grams category to 0.389 (0.175, 0.777) for the 22 weeks definition, suggesting that the 1000 grams definition data are on average 0.937 (0.929, 0.946) times lower than the 28 weeks definition, and 22 weeks definition data are 1.476 (1.192, 2.175) times higher than the 28 weeks definition data.
For low income countries, adjustments on the log-scale range from 0.214 (0.101, 0.426) for definitions in 22 weeks to 0.222 (0.058, 0.709) for definitions in 24 weeks, suggesting that 22 weeks definition data are on average 1.239 (1.106, 1.531) times higher than the 28 weeks definition, and 24 weeks definition data are on average 1.248 (1.060, 2.031) times higher than the 28 weeks definition.

\begin{table}[h]
\centering
\begin{tabular}{c|c|cc}
\hline
Definition & \begin{tabular}[c]{@{}c@{}}Income\\ group\end{tabular} & adjustment $\gamma_d$   & st.dev $\varphi_d$ \\ \hline
22 weeks      & High                                                   & 0.389  & 0.156  \\
22 weeks      & Low                                                    & 0.214  & 0.084  \\
24 weeks      & Low                                                    & 0.222  & 0.172  \\
1000 grams      & High                                                   & -0.065 & 0.073  \\
500 grams       & High                                                   & 0.244  & 0.087  \\ \hline
\end{tabular}
\caption{Adjustments and standard deviation of alternative definition versus the 28 week definition.}
\label{tb-def}
\end{table}

Table \ref{tab-source} summarizes the differences in error variance associated with the different source types, ranging from a standard deviation of 0.017 for national administrative data to 0.239 for review data. The bias for survey data is estimated at -0.165 (-0.229, -0.100) on the log-transformed scale, suggesting that survey data are on average 0.848 (0.795, 0.905) times lower than the truth. 

\begin{table}[h]
\centering
\begin{tabular}{ccc}
\hline
Source type       & Bias $\psi_j$ & st.dev $\sigma_j$ \\ \hline
Administrative    &  -   &  0.017      \\
HMIS              &  -   &  0.045      \\
Population study  &  -   &  0.239      \\
Survey            &  -0.165    & 0.135       \\ \hline
\end{tabular}
\caption{Source type bias and source type standard deviation.}
\label{tab-source}
\end{table}

\subsection{Illustrative findings}
Estimates for selected countries are given in Figure~\ref{fig:ill2}, with final estimates displayed in red and underlying covariate-based estimates in green. As highlighted earlier in the paper, data availability ranges in the selected countries from no data (Afghanistan) to an annual time series of national administrative data based on the 28 week definition for Ireland. The BHTSRM produces estimates for both countries. Estimates for Ireland are data-driven: point estimates track the observations closely and credible intervals are bounded by the uncertainty associated with each observation.
Estimates for Afghanistan are driven by covariates and the estimates are uncertain due to the absence of data.

Botswana, Malawi, Ukraine, and Uganda are examples of countries with SBR data that are either subject to bias, substantial error variance, or missing for periods of interest. In Ukraine, SBR data are available from 2007 to 2017 from administrative systems but recorded using a 22 week definition. SBR estimates are informed by the adjusted observations and uncertainty increases in extrapolations past the observation period. The survey data point has a large associated uncertainty and has little influence on the resulting model fit. In Uganda, the only available data come from HMIS, surveys, and population-based studies. There is substantial uncertainty associated with survey and population-based study data and resulting SBR estimates reflect this.  There are four different data sources in Botswana and Malawi. Resulting estimates are more certain in years with administrative or HMIS data as compared to population-based, survey, or no data. 

The effect of adding the smoother to the regression model on point estimates is visible in Ireland where final point estimates differ from the covariate-driven ones. In general, credible intervals are wider for the model that includes the smoother as shown in Figure~\ref{fig:ill2}. Exceptions include countries where data are limited except for a short period with low-variance data such as Malawi: in such countries, the addition of the smoother results in reduced uncertainty in the period with low-variance data (when the estimates are data-driven). 

\subsection{Covariates and subsetted model}\label{sec-res-cov}
Table~\ref{tab-cov} summarizes the estimates for regression coefficients.
As discussed in Section~\ref{sec-subset}, we check if we could obtain a more parsimonious model with similar estimates in order to produce the UN IGME stillbirth estimates. Based on the BHTSRM fit, we consider a subsetted model using an absolute cut off of 0.025, such that all covariates included in a prior study by \cite{blencowe_national_2016} were included, with the addition of one additional covariate. Comparisons between the model with horseshoe priors, the subsetted model, and an additional model are given in Appendix Section~\ref{sec-modelcompare}. In summary, there are no substantial differences among the models in terms of country estimates or model fit. Hence we used the subsetted model for producing UN IGME SBR estimates. Six covariates are included in the subsetted model: log-transformed NMR (nmr), log-transformed gross national income (gni), log-transformed percentage of live births with low birth weight (lbw), percentage of deliveries by Caesarian section (csec), average years of education received among women aged 25 and older (edu), and the percentage of women with at least four antenatal care visits during pregnancy (anc4). 

\begin{table}[ht]
\centering
\begin{tabular}{|l|c|c|c|c|c|c|}

\hline
	Covariates    & Estimate $\hat{\beta}$      & 2.5\% & 97.5\%  & SD(covariate)\\ \hline
log(nmr) & 0.414       & 0.336  & 0.492 & 0.999 \\ \hline
log(gni) & -0.102      & -0.212 & 0.001 & 1.20 \\ \hline
log(lbw) & 0.078       & 0.009  & 0.141 & 0.439 \\ \hline
edu      & -0.037      & -0.104 & 0.007 & 3.41 \\ \hline
csec     & -0.027      & -0.082 & 0.008 & 11.9 \\ \hline
anc4     & -0.025      & -0.094 & 0.014 & 21.8 \\ \hline
pab      & -0.018      & -0.050 & 0.006 & 11.6 \\ \hline
abr      & -0.017      & -0.109 & 0.023 & 46.5 \\ \hline
urban    & -0.012      & -0.087 & 0.024 & 23.1 \\ \hline
gini     & 0.010       & -0.017 & 0.061 & 8.17 \\ \hline
sab      & -0.010      & -0.083 & 0.026 & 0.215 \\ \hline
anc1     & -0.009      & -0.067 & 0.021 & 14.7\\ \hline
mmr      & 0.003       & -0.057 & 0.109 & 288.5\\ \hline
pfpr     & -0.002      & -0.045 & 0.030 & 0.118 \\ \hline
gdp      & 0.001       & -0.047 & 0.063 & $207 \cdot 10^2$\\ \hline
gfr      & 0.000       & -0.057 & 0.054 & 0.049 \\ \hline
 
  \end{tabular}
  \caption{Overview of estimates for regression coefficients under BHTSRM fit. Point estimates of regression coefficients, 95\% credible interval given by the 2.5th and 97.5th percentiles of the posterior, and the standard deviation of the covariate prior to standardization. Details on covariates are given in Appendix Table~\ref{tab-cov_inf}.}
\label{tab-cov}
\end{table}

\section{Discussion}\label{sec-discussion}
We develop a Bayesian hierarchical time series sparse regression model (BHTSRM) for estimating SBRs for all countries from 2000 until 2019. Estimating SBRs is challenging because of data paucity, especially for many LMIC where most stillbirths occur, and the substantial uncertainty associated with observations due to reporting issues and errors associated with the observations. Our BHTSRM extends the approach previously proposed by \cite{blencowe_national_2016} to produce estimates that are informed by a covariate model and available data, accounting for different definitions and uncertainty associated with the available data. Model validation exercises suggests that the model is reasonably well calibrated. 

The BHTSRM extends upon previous applications of Bayesian hierarchical time series regression models through the introduction of sparsity-inducing priors and new statistical approaches to addresses data quality issues. Sparsity-inducing priors allow for the inclusion of larger sets of (potentially correlated) candidate covariates into the model. To address data quality issues, we develop a procedure for data exclusion based on comparing observed ratios of SBR to NMR for the population of interest to a reference distribution of such ratios. Secondly, we develop a new approach to adjust and estimate additional uncertainty associated with  observations using a different definition of stillbirths. In the model fitting, we use a data model that accounts for bias and varying sources of random error associated with the observations.

While our approach to estimating the SBR improves upon existing approaches, there are limitations related to the model and data availability. Limited data availability restricted the analyses we are able to carry out and result in stricter modeling assumptions. For example, relative differences in SBRs associated with the use of different definitions, i.e, gestational age, may vary across settings. Data limitations result in the use of a simple dichotomy of high income and low income countries to capture this difference. With additional data, this relationship can be studied in more detail. 

The BHTSRM as described in this paper is used by the UN IGME to generate estimates for the SBR globally.  While the modeling approach allows for the construction of estimates for all countries, we find that uncertainty associated with the estimates is substantial in many settings,  including countries with high SBRs. This highlights the need for additional data collection to produce more precise information for monitoring and program planning, especially in high-burden settings.

\section*{Acknowledgements}

We thank all members of the Core Stillbirth Estimation Group (Leontine Alkema, Dianna Blau, Simon Cousens, Andreea Creanga, Trevor Croft, KS Joseph, Salome Maswime, Elizabeth McClure, Robert Pattinson, Jon Pedersen, Lucy Smith, Jennifer Zeitlin) and the members of the Technical Advisory Group of the UN Inter-agency Group for Child Mortality Estimation (Leontine Alkema, Robert Black, Simon Cousens, Trevor Croft, Michel Guillot, Kenneth Hill, Bruno Masquelier, Colin Mathers, Jon Pedersen, Jon Wakefield, Neff Walker) for providing input and feedback on estimating stillbirths. We thank the members of the UN Inter-agency Group for Child Mortality Estimation (Allisyn Moran, Emi Suzuki, Victor Gaigbe-Togbe) for additional guidance. We also thank Remy Wang, Serena Wang, and Zitong Wang for their contributions to the modeling in an earlier stage of the project. 

\bibliographystyle{imsart-nameyear}
\bibliography{AlkemaLab}

\begin{thebibliography}{27}

\bibitem[\protect\citeauthoryear{Ahmed
  et~al.}{2018}]{ahmed_population-based_2018}
\begin{barticle}[author]
\bauthor{\bsnm{Ahmed},~\bfnm{Imran}\binits{I.}},
  \bauthor{\bsnm{Ali},~\bfnm{Said~Mohammed}\binits{S.~M.}},
  \bauthor{\bsnm{Amenga-Etego},~\bfnm{Seeba}\binits{S.}},
  \bauthor{\bsnm{Ariff},~\bfnm{Shabina}\binits{S.}},
  \bauthor{\bsnm{Bahl},~\bfnm{Rajiv}\binits{R.}},
  \bauthor{\bsnm{Baqui},~\bfnm{Abdullah~H}\binits{A.~H.}},
  \bauthor{\bsnm{Begum},~\bfnm{Nazma}\binits{N.}},
  \bauthor{\bsnm{Bhandari},~\bfnm{Nita}\binits{N.}},
  \bauthor{\bsnm{Bhatia},~\bfnm{Kiran}\binits{K.}},
  \bauthor{\bsnm{Bhutta},~\bfnm{Zulfiqar~A}\binits{Z.~A.}},
  \bauthor{\bsnm{Biemba},~\bfnm{Godfrey}\binits{G.}},
  \bauthor{\bsnm{Deb},~\bfnm{Saikat}\binits{S.}},
  \bauthor{\bsnm{Dhingra},~\bfnm{Usha}\binits{U.}},
  \bauthor{\bsnm{Dube},~\bfnm{Brinda}\binits{B.}},
  \bauthor{\bsnm{Dutta},~\bfnm{Arup}\binits{A.}},
  \bauthor{\bsnm{Edmond},~\bfnm{Karen}\binits{K.}},
  \bauthor{\bsnm{Esamai},~\bfnm{Fabian}\binits{F.}},
  \bauthor{\bsnm{Fawzi},~\bfnm{Wafaie}\binits{W.}},
  \bauthor{\bsnm{Ghosh},~\bfnm{Amit~Kumar}\binits{A.~K.}},
  \bauthor{\bsnm{Gisore},~\bfnm{Peter}\binits{P.}},
  \bauthor{\bsnm{Grogan},~\bfnm{Caroline}\binits{C.}},
  \bauthor{\bsnm{Hamer},~\bfnm{Davidson~H}\binits{D.~H.}},
  \bauthor{\bsnm{Herlihy},~\bfnm{Julie}\binits{J.}},
  \bauthor{\bsnm{Hurt},~\bfnm{Lisa}\binits{L.}},
  \bauthor{\bsnm{Ilyas},~\bfnm{Muhammad}\binits{M.}},
  \bauthor{\bsnm{Jehan},~\bfnm{Fyezah}\binits{F.}},
  \bauthor{\bsnm{Kalonji},~\bfnm{Michel}\binits{M.}},
  \bauthor{\bsnm{Kaur},~\bfnm{Jasmine}\binits{J.}},
  \bauthor{\bsnm{Khanam},~\bfnm{Rasheda}\binits{R.}},
  \bauthor{\bsnm{Kirkwood},~\bfnm{Betty}\binits{B.}},
  \bauthor{\bsnm{Kumar},~\bfnm{Aarti}\binits{A.}},
  \bauthor{\bsnm{Kumar},~\bfnm{Alok}\binits{A.}},
  \bauthor{\bsnm{Kumar},~\bfnm{Vishwajeet}\binits{V.}},
  \bauthor{\bsnm{Manu},~\bfnm{Alexander}\binits{A.}},
  \bauthor{\bsnm{Marete},~\bfnm{Irene}\binits{I.}},
  \bauthor{\bsnm{Masanja},~\bfnm{Honorati}\binits{H.}},
  \bauthor{\bsnm{Mazumder},~\bfnm{Sarmila}\binits{S.}},
  \bauthor{\bsnm{Mehmood},~\bfnm{Usma}\binits{U.}},
  \bauthor{\bsnm{Mishra},~\bfnm{Shambhavi}\binits{S.}},
  \bauthor{\bsnm{Mitra},~\bfnm{Dipak~K}\binits{D.~K.}},
  \bauthor{\bsnm{Mlay},~\bfnm{Erick}\binits{E.}},
  \bauthor{\bsnm{Mohan},~\bfnm{Sanjana~Brahmawar}\binits{S.~B.}},
  \bauthor{\bsnm{Moin},~\bfnm{Mamun~Ibne}\binits{M.~I.}},
  \bauthor{\bsnm{Muhammad},~\bfnm{Karim}\binits{K.}},
  \bauthor{\bsnm{Muhihi},~\bfnm{Alfa}\binits{A.}},
  \bauthor{\bsnm{Newton},~\bfnm{Samuel}\binits{S.}},
  \bauthor{\bsnm{Ngaima},~\bfnm{Serge}\binits{S.}},
  \bauthor{\bsnm{Nguwo},~\bfnm{Andre}\binits{A.}},
  \bauthor{\bsnm{Nisar},~\bfnm{Imran}\binits{I.}},
  \bauthor{\bsnm{O'Leary},~\bfnm{Maureen}\binits{M.}},
  \bauthor{\bsnm{Otomba},~\bfnm{John}\binits{J.}},
  \bauthor{\bsnm{Patil},~\bfnm{Pawankumar}\binits{P.}},
  \bauthor{\bsnm{Quaiyum},~\bfnm{Mohammad~Abdul}\binits{M.~A.}},
  \bauthor{\bsnm{Rahman},~\bfnm{Mohammed~Hefzur}\binits{M.~H.}},
  \bauthor{\bsnm{Sazawal},~\bfnm{Sunil}\binits{S.}},
  \bauthor{\bsnm{Semrau},~\bfnm{Katherine~EA}\binits{K.~E.}},
  \bauthor{\bsnm{Shannon},~\bfnm{Caitlin}\binits{C.}},
  \bauthor{\bsnm{Smith},~\bfnm{Emily~R}\binits{E.~R.}},
  \bauthor{\bsnm{Soofi},~\bfnm{Sajid}\binits{S.}},
  \bauthor{\bsnm{Soremekun},~\bfnm{Seyi}\binits{S.}},
  \bauthor{\bsnm{Sunday},~\bfnm{Venantius}\binits{V.}},
  \bauthor{\bsnm{Taneja},~\bfnm{Sunita}\binits{S.}},
  \bauthor{\bsnm{Tshefu},~\bfnm{Antoinette}\binits{A.}},
  \bauthor{\bsnm{Wasan},~\bfnm{Yaqub}\binits{Y.}},
  \bauthor{\bsnm{Yeboah-Antwi},~\bfnm{Kojo}\binits{K.}},
  \bauthor{\bsnm{Yoshida},~\bfnm{Sachiyo}\binits{S.}} \AND
  \bauthor{\bsnm{Zaidi},~\bfnm{Anita}\binits{A.}}
(\byear{2018}).
\btitle{Population-based rates, timing, and causes of maternal deaths,
  stillbirths, and neonatal deaths in south {Asia} and sub-{Saharan} {Africa}:
  a multi-country prospective cohort study}.
\bjournal{The Lancet Global Health}
\bvolume{6}
\bpages{e1297--e1308}.
\bdoi{10.1016/S2214-109X(18)30385-1}
\end{barticle}
\endbibitem

\bibitem[\protect\citeauthoryear{Alkema et~al.}{2017}]{alkema_bayesian_2017}
\begin{barticle}[author]
\bauthor{\bsnm{Alkema},~\bfnm{Leontine}\binits{L.}},
  \bauthor{\bsnm{Zhang},~\bfnm{Sanqian}\binits{S.}},
  \bauthor{\bsnm{Chou},~\bfnm{Doris}\binits{D.}},
  \bauthor{\bsnm{Gemmill},~\bfnm{Alison}\binits{A.}},
  \bauthor{\bsnm{Moller},~\bfnm{Ann-Beth}\binits{A.-B.}},
  \bauthor{\bsnm{Fat},~\bfnm{Doris~Ma}\binits{D.~M.}},
  \bauthor{\bsnm{Say},~\bfnm{Lale}\binits{L.}},
  \bauthor{\bsnm{Mathers},~\bfnm{Colin}\binits{C.}} \AND
  \bauthor{\bsnm{Hogan},~\bfnm{Daniel}\binits{D.}}
(\byear{2017}).
\btitle{A {Bayesian} approach to the global estimation of maternal mortality}.
\bjournal{The Annals of Applied Statistics}
\bvolume{11}
\bpages{1245--1274}.
\bdoi{10.1214/16-AOAS1014}
\end{barticle}
\endbibitem

\bibitem[\protect\citeauthoryear{Blencowe
  et~al.}{2016}]{blencowe_national_2016}
\begin{barticle}[author]
\bauthor{\bsnm{Blencowe},~\bfnm{Hannah}\binits{H.}},
  \bauthor{\bsnm{Cousens},~\bfnm{Simon}\binits{S.}},
  \bauthor{\bsnm{Jassir},~\bfnm{Fiorella~Bianchi}\binits{F.~B.}},
  \bauthor{\bsnm{Say},~\bfnm{Lale}\binits{L.}},
  \bauthor{\bsnm{Chou},~\bfnm{Doris}\binits{D.}},
  \bauthor{\bsnm{Mathers},~\bfnm{Colin}\binits{C.}},
  \bauthor{\bsnm{Hogan},~\bfnm{Dan}\binits{D.}},
  \bauthor{\bsnm{Shiekh},~\bfnm{Suhail}\binits{S.}},
  \bauthor{\bsnm{Qureshi},~\bfnm{Zeshan~U}\binits{Z.~U.}},
  \bauthor{\bsnm{You},~\bfnm{Danzhen}\binits{D.}} \AND
  \bauthor{\bsnm{Lawn},~\bfnm{Joy~E}\binits{J.~E.}}
(\byear{2016}).
\btitle{National, regional, and worldwide estimates of stillbirth rates in
  2015, with trends from 2000: a systematic analysis}.
\bjournal{The Lancet Global Health}
\bvolume{4}
\bpages{e98--e108}.
\bdoi{10.1016/S2214-109X(15)00275-2}
\end{barticle}
\endbibitem

\bibitem[\protect\citeauthoryear{Bose et~al.}{2015}]{bose_global_2015}
\begin{barticle}[author]
\bauthor{\bsnm{Bose},~\bfnm{Carl~L}\binits{C.~L.}},
  \bauthor{\bsnm{Bauserman},~\bfnm{Melissa}\binits{M.}},
  \bauthor{\bsnm{Goldenberg},~\bfnm{Robert~L}\binits{R.~L.}},
  \bauthor{\bsnm{Goudar},~\bfnm{Shivaprasad~S}\binits{S.~S.}},
  \bauthor{\bsnm{McClure},~\bfnm{Elizabeth~M}\binits{E.~M.}},
  \bauthor{\bsnm{Pasha},~\bfnm{Omrana}\binits{O.}},
  \bauthor{\bsnm{Carlo},~\bfnm{Waldemar~A}\binits{W.~A.}},
  \bauthor{\bsnm{Garces},~\bfnm{Ana}\binits{A.}},
  \bauthor{\bsnm{Moore},~\bfnm{Janet~L}\binits{J.~L.}},
  \bauthor{\bsnm{Miodovnik},~\bfnm{Menachem}\binits{M.}} \AND
  \bauthor{\bsnm{Koso-Thomas},~\bfnm{Marion}\binits{M.}}
(\byear{2015}).
\btitle{The {Global} {Network} {Maternal} {Newborn} {Health} {Registry}: a
  multi-national, community-based registry of pregnancy outcomes}.
\bjournal{Reproductive Health}
\bvolume{12}
\bpages{S1}.
\bdoi{10.1186/1742-4755-12-S2-S1}
\end{barticle}
\endbibitem

\bibitem[\protect\citeauthoryear{Bradley, Winfrey and
  Croft}{2015}]{Bradley_contraceptive_2015}
\begin{barticle}[author]
\bauthor{\bsnm{Bradley},~\bfnm{Sarah E.~K.}\binits{S.~E.~K.}},
  \bauthor{\bsnm{Winfrey},~\bfnm{William}\binits{W.}} \AND
  \bauthor{\bsnm{Croft},~\bfnm{Trevor~N.}\binits{T.~N.}}
(\byear{2015}).
\btitle{Contraceptive Use and Perinatal Mortality in the DHS: An Assessment of
  the Quality and Consistency of Calendars and Histories.}
\bnote{DHS Methodological Reports No. 17. Rockville, Maryland, USA: ICF
  International.}
\end{barticle}
\endbibitem

\bibitem[\protect\citeauthoryear{Carpenter et~al.}{2017}]{carpenter_stan_2017}
\begin{barticle}[author]
\bauthor{\bsnm{Carpenter},~\bfnm{Bob}\binits{B.}},
  \bauthor{\bsnm{Gelman},~\bfnm{Andrew}\binits{A.}},
  \bauthor{\bsnm{Hoffman},~\bfnm{Matthew~D.}\binits{M.~D.}},
  \bauthor{\bsnm{Lee},~\bfnm{Daniel}\binits{D.}},
  \bauthor{\bsnm{Goodrich},~\bfnm{Ben}\binits{B.}},
  \bauthor{\bsnm{Betancourt},~\bfnm{Michael}\binits{M.}},
  \bauthor{\bsnm{Brubaker},~\bfnm{Marcus}\binits{M.}},
  \bauthor{\bsnm{Guo},~\bfnm{Jiqiang}\binits{J.}},
  \bauthor{\bsnm{Li},~\bfnm{Peter}\binits{P.}} \AND
  \bauthor{\bsnm{Riddell},~\bfnm{Allen}\binits{A.}}
(\byear{2017}).
\btitle{\textit{{Stan}} : {A} {Probabilistic} {Programming} {Language}}.
\bjournal{Journal of Statistical Software}
\bvolume{76}.
\bdoi{10.18637/jss.v076.i01}
\end{barticle}
\endbibitem

\bibitem[\protect\citeauthoryear{Carvalho, Polson and
  Scott}{2009}]{carvalho_handling_nodate}
\begin{barticle}[author]
\bauthor{\bsnm{Carvalho},~\bfnm{Carlos~M}\binits{C.~M.}},
  \bauthor{\bsnm{Polson},~\bfnm{Nicholas~G}\binits{N.~G.}} \AND
  \bauthor{\bsnm{Scott},~\bfnm{James~G}\binits{J.~G.}}
(\byear{2009}).
\btitle{Handling {Sparsity} via the {Horseshoe}}.
\bjournal{Artificial Intelligence and Statistics}
\bpages{8}.
\end{barticle}
\endbibitem

\bibitem[\protect\citeauthoryear{Christou et~al.}{2019}]{christou_how_2019}
\begin{barticle}[author]
\bauthor{\bsnm{Christou},~\bfnm{Aliki}\binits{A.}},
  \bauthor{\bsnm{Alam},~\bfnm{Ashraful}\binits{A.}},
  \bauthor{\bsnm{Hofiani},~\bfnm{Sayed Murtaza~Sadat}\binits{S.~M.~S.}},
  \bauthor{\bsnm{Rasooly},~\bfnm{Mohammad~Hafiz}\binits{M.~H.}},
  \bauthor{\bsnm{Mubasher},~\bfnm{Adela}\binits{A.}},
  \bauthor{\bsnm{Rashidi},~\bfnm{Mohammad~Khakerah}\binits{M.~K.}},
  \bauthor{\bsnm{Dibley},~\bfnm{Michael~J.}\binits{M.~J.}} \AND
  \bauthor{\bsnm{Raynes-Greenow},~\bfnm{Camille}\binits{C.}}
(\byear{2019}).
\btitle{How community and healthcare provider perceptions, practices and
  experiences influence reporting, disclosure and data collection on
  stillbirth: {Findings} of a qualitative study in {Afghanistan}}.
\bjournal{Social Science \& Medicine}
\bvolume{236}
\bpages{112413}.
\bdoi{10.1016/j.socscimed.2019.112413}
\end{barticle}
\endbibitem

\bibitem[\protect\citeauthoryear{Currie and
  Durban}{2002}]{currie_flexible_2002}
\begin{barticle}[author]
\bauthor{\bsnm{Currie},~\bfnm{I~D}\binits{I.~D.}} \AND
  \bauthor{\bsnm{Durban},~\bfnm{M}\binits{M.}}
(\byear{2002}).
\btitle{Flexible smoothing with {P}-splines: a unified approach}.
\bjournal{Statistical Modelling: An International Journal}
\bvolume{2}
\bpages{333--349}.
\bdoi{10.1191/1471082x02st039ob}
\end{barticle}
\endbibitem

\bibitem[\protect\citeauthoryear{Eilers and Marx}{1996}]{eilers_flexible_1996}
\begin{barticle}[author]
\bauthor{\bsnm{Eilers},~\bfnm{Paul H.~C.}\binits{P.~H.~C.}} \AND
  \bauthor{\bsnm{Marx},~\bfnm{Brian~D.}\binits{B.~D.}}
(\byear{1996}).
\btitle{Flexible smoothing with {B} -splines and penalties}.
\bjournal{Statistical Science}
\bvolume{11}
\bpages{89--121}.
\bdoi{10.1214/ss/1038425655}
\end{barticle}
\endbibitem

\bibitem[\protect\citeauthoryear{Gelman}{2006}]{gelman_prior_2006}
\begin{barticle}[author]
\bauthor{\bsnm{Gelman},~\bfnm{Andrew}\binits{A.}}
(\byear{2006}).
\btitle{Prior distributions for variance parameters in hierarchical models
  (comment on article by {Browne} and {Draper})}.
\bjournal{Bayesian Analysis}
\bvolume{1}
\bpages{515--534}.
\bdoi{10.1214/06-BA117A}
\end{barticle}
\endbibitem

\bibitem[\protect\citeauthoryear{Gelman and Rubin}{1992}]{mcmc_diag_GR}
\begin{barticle}[author]
\bauthor{\bsnm{Gelman},~\bfnm{Andrew}\binits{A.}} \AND
  \bauthor{\bsnm{Rubin},~\bfnm{B.}\binits{B.}}
(\byear{1992}).
\btitle{Inference from iterative simulation using multiple sequences}.
\bjournal{Statistical Science}
\bvolume{7}
\bpages{457–72}.
\end{barticle}
\endbibitem

\bibitem[\protect\citeauthoryear{Kuruvilla
  et~al.}{2016}]{kuruvilla_global_2016}
\begin{barticle}[author]
\bauthor{\bsnm{Kuruvilla},~\bfnm{Shyama}\binits{S.}},
  \bauthor{\bsnm{Bustreo},~\bfnm{Flavia}\binits{F.}},
  \bauthor{\bsnm{Kuo},~\bfnm{Taona}\binits{T.}},
  \bauthor{\bsnm{Mishra},~\bfnm{Ck}\binits{C.}},
  \bauthor{\bsnm{Taylor},~\bfnm{Katie}\binits{K.}},
  \bauthor{\bsnm{Fogstad},~\bfnm{Helga}\binits{H.}},
  \bauthor{\bsnm{Gupta},~\bfnm{Geeta~Rao}\binits{G.~R.}},
  \bauthor{\bsnm{Gilmore},~\bfnm{Kate}\binits{K.}},
  \bauthor{\bsnm{Temmerman},~\bfnm{Marleen}\binits{M.}},
  \bauthor{\bsnm{Thomas},~\bfnm{Joe}\binits{J.}},
  \bauthor{\bsnm{Rasanathan},~\bfnm{Kumanan}\binits{K.}},
  \bauthor{\bsnm{Chaiban},~\bfnm{Ted}\binits{T.}},
  \bauthor{\bsnm{Mohan},~\bfnm{Anshu}\binits{A.}},
  \bauthor{\bsnm{Gruending},~\bfnm{Anna}\binits{A.}},
  \bauthor{\bsnm{Schweitzer},~\bfnm{Julian}\binits{J.}},
  \bauthor{\bsnm{Dini},~\bfnm{Hannah~Sarah}\binits{H.~S.}},
  \bauthor{\bsnm{Borrazzo},~\bfnm{John}\binits{J.}},
  \bauthor{\bsnm{Fassil},~\bfnm{Hareya}\binits{H.}},
  \bauthor{\bsnm{Gronseth},~\bfnm{Lars}\binits{L.}},
  \bauthor{\bsnm{Khosla},~\bfnm{Rajat}\binits{R.}},
  \bauthor{\bsnm{Cheeseman},~\bfnm{Richard}\binits{R.}},
  \bauthor{\bsnm{Gorna},~\bfnm{Robin}\binits{R.}},
  \bauthor{\bsnm{McDougall},~\bfnm{Lori}\binits{L.}},
  \bauthor{\bsnm{Toure},~\bfnm{Kadidiatou}\binits{K.}},
  \bauthor{\bsnm{Rogers},~\bfnm{Kate}\binits{K.}},
  \bauthor{\bsnm{Dodson},~\bfnm{Kate}\binits{K.}},
  \bauthor{\bsnm{Sharma},~\bfnm{Anita}\binits{A.}},
  \bauthor{\bsnm{Seoane},~\bfnm{Marta}\binits{M.}} \AND
  \bauthor{\bsnm{Costello},~\bfnm{Anthony}\binits{A.}}
(\byear{2016}).
\btitle{The \textit{{Global} strategy for women’s, children’s and
  adolescents’ health (2016–2030)} : a roadmap based on evidence and
  country experience}.
\bjournal{Bulletin of the World Health Organization}
\bvolume{94}
\bpages{398--400}.
\bdoi{10.2471/BLT.16.170431}
\end{barticle}
\endbibitem

\bibitem[\protect\citeauthoryear{Liu et~al.}{2016}]{liu_understanding_2016}
\begin{barticle}[author]
\bauthor{\bsnm{Liu},~\bfnm{Li}\binits{L.}},
  \bauthor{\bsnm{Kalter},~\bfnm{Henry~D.}\binits{H.~D.}},
  \bauthor{\bsnm{Chu},~\bfnm{Yue}\binits{Y.}},
  \bauthor{\bsnm{Kazmi},~\bfnm{Narjis}\binits{N.}},
  \bauthor{\bsnm{Koffi},~\bfnm{Alain~K.}\binits{A.~K.}},
  \bauthor{\bsnm{Amouzou},~\bfnm{Agbessi}\binits{A.}},
  \bauthor{\bsnm{Joos},~\bfnm{Olga}\binits{O.}},
  \bauthor{\bsnm{Munos},~\bfnm{Melinda}\binits{M.}} \AND
  \bauthor{\bsnm{Black},~\bfnm{Robert~E.}\binits{R.~E.}}
(\byear{2016}).
\btitle{Understanding {Misclassification} between {Neonatal} {Deaths} and
  {Stillbirths}: {Empirical} {Evidence} from {Malawi}}.
\bjournal{PLOS ONE}
\bvolume{11}
\bpages{e0168743}.
\bdoi{10.1371/journal.pone.0168743}
\end{barticle}
\endbibitem

\bibitem[\protect\citeauthoryear{{World Health
  Organization}}{2014}]{world_health_organization_every_2014}
\begin{bbook}[author]
\bauthor{\bsnm{{World Health Organization}}}
(\byear{2014}).
\btitle{Every newborn: an action plan to end preventable deaths}.
\bpublisher{World Health Organization}, \baddress{Geneva}.
\bnote{OCLC: 918973731}.
\end{bbook}
\endbibitem

\bibitem[\protect\citeauthoryear{{World Health Organization}}{2019}]{icd11}
\begin{bbook}[author]
\bauthor{\bsnm{{World Health Organization}}}
(\byear{2019}).
\btitle{International statistical classification of diseases and related health
  problems, tenth revision: Instruction manual.(11th revision)}.
\bpublisher{World Health Organization}.
\end{bbook}
\endbibitem

\bibitem[\protect\citeauthoryear{Pedersen}{2012}]{CMRjack}
\begin{bmanual}[author]
\bauthor{\bsnm{Pedersen},~\bfnm{Jon}\binits{J.}}
(\byear{2012}).
\btitle{CMRjack}
\bnote{\url{https://cmrjack.org/}}.
\end{bmanual}
\endbibitem

\bibitem[\protect\citeauthoryear{Piironen and
  Vehtari}{2017a}]{piironen_sparsity_2017}
\begin{barticle}[author]
\bauthor{\bsnm{Piironen},~\bfnm{Juho}\binits{J.}} \AND
  \bauthor{\bsnm{Vehtari},~\bfnm{Aki}\binits{A.}}
(\byear{2017}a).
\btitle{Sparsity information and regularization in the horseshoe and other
  shrinkage priors}.
\bjournal{Electronic Journal of Statistics}
\bvolume{11}
\bpages{5018--5051}.
\bdoi{10.1214/17-EJS1337SI}
\end{barticle}
\endbibitem

\bibitem[\protect\citeauthoryear{Piironen and
  Vehtari}{2017b}]{piironen_comparison_2017}
\begin{barticle}[author]
\bauthor{\bsnm{Piironen},~\bfnm{Juho}\binits{J.}} \AND
  \bauthor{\bsnm{Vehtari},~\bfnm{Aki}\binits{A.}}
(\byear{2017}b).
\btitle{Comparison of {Bayesian} predictive methods for model selection}.
\bjournal{Statistics and Computing}
\bvolume{27}
\bpages{711--735}.
\bdoi{10.1007/s11222-016-9649-y}
\end{barticle}
\endbibitem

\bibitem[\protect\citeauthoryear{Stanton
  et~al.}{2006}]{stanton_stillbirth_2006}
\begin{barticle}[author]
\bauthor{\bsnm{Stanton},~\bfnm{Cynthia}\binits{C.}},
  \bauthor{\bsnm{Lawn},~\bfnm{Joy~E}\binits{J.~E.}},
  \bauthor{\bsnm{Rahman},~\bfnm{Hafiz}\binits{H.}},
  \bauthor{\bsnm{Wilczynska-Ketende},~\bfnm{Katarzyna}\binits{K.}} \AND
  \bauthor{\bsnm{Hill},~\bfnm{Kenneth}\binits{K.}}
(\byear{2006}).
\btitle{Stillbirth rates: delivering estimates in 190 countries}.
\bjournal{The Lancet}
\bvolume{367}
\bpages{1487--1494}.
\bdoi{10.1016/S0140-6736(06)68586-3}
\end{barticle}
\endbibitem

\bibitem[\protect\citeauthoryear{{Stan Development Team}}{2018}]{rstan_package}
\begin{bmisc}[author]
\bauthor{\bsnm{{Stan Development Team}}}
(\byear{2018}).
\btitle{{RStan}: the {R} interface to {Stan}}.
\bnote{R package version 2.18.2}.
\end{bmisc}
\endbibitem

\bibitem[\protect\citeauthoryear{{Stan Development Team}}{2020}]{STANwarnings}
\begin{bmanual}[author]
\bauthor{\bsnm{{Stan Development Team}}}
(\byear{2020}).
\btitle{Brief Guide to Stan's Warnings}
\bnote{\url{https://mc-stan.org/misc/warnings.html}}.
\end{bmanual}
\endbibitem

\bibitem[\protect\citeauthoryear{{United Nations, Department of Economic and
  Social Affairs, Population Division}}{2019}]{wpp2019}
\begin{bbook}[author]
\bauthor{\bsnm{{United Nations, Department of Economic and Social Affairs,
  Population Division}}}
(\byear{2019}).
\btitle{World Population Prospects. The 2019 revision.}
\end{bbook}
\endbibitem

\bibitem[\protect\citeauthoryear{Vehtari, Gelman and
  Gabry}{2017}]{vehtari_practical_2017}
\begin{barticle}[author]
\bauthor{\bsnm{Vehtari},~\bfnm{Aki}\binits{A.}},
  \bauthor{\bsnm{Gelman},~\bfnm{Andrew}\binits{A.}} \AND
  \bauthor{\bsnm{Gabry},~\bfnm{Jonah}\binits{J.}}
(\byear{2017}).
\btitle{Practical {Bayesian} model evaluation using leave-one-out
  cross-validation and {WAIC}}.
\bjournal{Statistics and Computing}
\bvolume{27}
\bpages{1413--1432}.
\bdoi{10.1007/s11222-016-9696-4}
\end{barticle}
\endbibitem

\bibitem[\protect\citeauthoryear{Vehtari et~al.}{2019}]{loo_package}
\begin{bmisc}[author]
\bauthor{\bsnm{Vehtari},~\bfnm{Aki}\binits{A.}},
  \bauthor{\bsnm{Gabry},~\bfnm{Jonah}\binits{J.}},
  \bauthor{\bsnm{Magnusson},~\bfnm{Mans}\binits{M.}},
  \bauthor{\bsnm{Yao},~\bfnm{Yuling}\binits{Y.}} \AND
  \bauthor{\bsnm{Gelman},~\bfnm{Andrew}\binits{A.}}
(\byear{2019}).
\btitle{loo: Efficient leave-one-out cross-validation and WAIC for Bayesian
  models}.
\bnote{R package version 2.2.0}.
\end{bmisc}
\endbibitem

\bibitem[\protect\citeauthoryear{Vehtari
  et~al.}{2020}]{vehtari_rank-normalization_2020}
\begin{barticle}[author]
\bauthor{\bsnm{Vehtari},~\bfnm{Aki}\binits{A.}},
  \bauthor{\bsnm{Gelman},~\bfnm{Andrew}\binits{A.}},
  \bauthor{\bsnm{Simpson},~\bfnm{Daniel}\binits{D.}},
  \bauthor{\bsnm{Carpenter},~\bfnm{Bob}\binits{B.}} \AND
  \bauthor{\bsnm{Bürkner},~\bfnm{Paul-Christian}\binits{P.-C.}}
(\byear{2020}).
\btitle{Rank-{Normalization}, {Folding}, and {Localization}: {An} {Improved} {R
  hat} for {Assessing} {Convergence} of {MCMC}}.
\bjournal{Bayesian Analysis}.
\bdoi{10.1214/20-BA1221}
\end{barticle}
\endbibitem

\bibitem[\protect\citeauthoryear{Woods}{2008}]{woods_long-term_2008}
\begin{barticle}[author]
\bauthor{\bsnm{Woods},~\bfnm{Robert}\binits{R.}}
(\byear{2008}).
\btitle{Long-term tends in fetal mortality: implications for developing
  countries}.
\bjournal{Bulletin of the World Health Organization}
\bvolume{2008}
\bpages{460--466}.
\bdoi{10.2471/BLT.07.043471}
\end{barticle}
\endbibitem

\end{thebibliography}

\section{Appendix}

\subsection{Data processing}\label{sec-dataexclusion}

The process of compiling and processing the stillbirth database for analysis was composed of two steps:
\begin{enumerate}
\item	Compile all available stillbirth data at a country level from 2000 onward, derived from administrative sources, HMIS, household surveys or population-based studies.
\item	Evaluate data in accordance with the data quality criteria and produce adjustment or recalculation by applying standardized definitions. 
\end{enumerate}

The majority of data collected on stillbirths were obtained from administrative data systems and health data systems including health management information systems (HMIS). UN IGME conducts an annual country consultation to solicit up to date administrative data on stillbirths from ministries of health or national statistics offices. After data were compiled, stillbirth data were processed. In some instances, proportions of administrative stillbirth data had unknown gestational age or birthweight. We excluded observations for which the proportion of reported stillbirths with unknown gestational age or birthweight was above 50 per cent. For other data, stillbirths with unknown gestational age or birthweight were redistributed proportionally among the distribution of stillbirths from that same country-year. 

Nationally representative household surveys (e.g. Demographic and Health Surveys, Multiple Indicator Cluster Surveys, Reproductive Health surveys) are another source of stillbirth data obtained from household survey data. Information on stillbirths in household surveys can be collected with a full pregnancy history (PH) or with a reproductive calendar (RC). Stillbirth estimates from the RC were not included in the model if estimates from the PH in the same survey were available. In pregnancy histories, the SBR is the number of stillbirths with the end of the pregnancy in the seventh month or later divided by the number of stillbirths plus livebirths. In some surveys with PH modules, the women were only asked whether they had a stillbirth and the date of the stillbirth. In these cases, a seven month duration of pregnancy was assumed. In some survey-specific cases, a stillbirth was defined by the questionnaire as a fetal death occurring at the fifth or sixth month or later. In RCs, the SBR is the number of pregnancies that are terminated in the seventh month or later of pregnancy divided by the number of pregnancies that reached at least the seventh month. Stillbirth estimates were calculated for the most recent 5-year period prior to the survey. Where the microdata were available, stillbirth estimates were recalculated with standard errors. 

Stillbirth data from subnational population-based studies were obtained via literature reviews. The literature review undertaken for the previous stillbirth estimates (\cite{blencowe_national_2016}) was updated through to 29 January 2019. In addition, further reanalyzed population-based stillbirth data were obtained from a WHO data call to maternal-newborn health experts.

The evaluation and assessment for data quality were applied to all data sources based on pre-defined exclusion criteria. Data were excluded if they lacked information on definition or data collection systems, if the proportion of reported stillbirths with unknown gestational age or birthweight was above 50 per cent, if data were internally inconsistent, or if coverage of live births in administrative data systems was estimated below 80 per cent. Vital registration data with incomplete coverage of child deaths were also excluded, where incompleteness was taken from the WHO CRVS completeness assessment.

\subsection{Variance of the ratio of SBR to NMR} \label{sec-app-ex}
The variance term $v_i^2$ in Equation~(\ref{eq-ratio2}) refers to the variance of the random error associated with the $i$-th observed log-transformed SBR to NMR ratio. We use a Monte Carlo approximation to calculate the observation-specific error variance $v_i^2$. We denote $z_i$ as stillbirths and $m_i$ as neonatal deaths, and assume:

\begin{eqnarray*}
z_i|p_i^{(sb)} &\sim& Bin(t_i,p_i^{(sb)}),\\
m_i|p_i^{(nd)} &\sim& Bin(q_i,p_i^{(nd)}),
\end{eqnarray*}

where $t_i$ and $p_i^{(sb)}$ refer to total births and observed SBR and $q_i$ and $p_i^{(nd)}$ refer to the number of live births and NMR. Assuming independence between stillbirths and neonatal deaths, we obtain samples $(z_i^{(s)}, m_i^{(s)})$ and calculate the associated ratio $r_i^{(s)}$:
\begin{eqnarray*}
r_i^{(s)} = \frac{z_i^{(s)}/t_i}{m_i^{(s)}/q_i}.
\end{eqnarray*}
Variance $v_i^2$ is given by the variance of the samples $\log(r_i^{(s)})$.

\subsection{Variance of SBR}\label{sec-se}
The variance term $s_i^2$ refers to the variance of observed log SBR in Equation~(\ref{eq-dm}). For observations from surveys, sampling error $s_i$ is pre-calculated using a jackknife method (\cite{CMRjack}), to reflect the survey sampling design. 
For observations from registration systems, we assume a Poisson data-generating process to obtain $s_i^2$. Specifically, for SBR rate $y = D/B$, with deaths $D$ and total births $B$, we assume $D\mid \Theta \sim Poisson(B\times \Theta)$. Then $\textit{var}(y)=D/B^2$ and by using the delta method, we obtain: 
\begin{equation*}
    \textit{var}(\log(y))  = \frac{1}{B\times y}.
\end{equation*}
Therefore, the variance $s_i^2$ for the $i$-th observation is set to  $ \frac{1}{B_iy_i}$, where $B_i$ and $y_i$ are the number of total birth and observed SBR rate for the $i$-th observation, respectively. Total births are obtained from the data source where available. If unknown, births are calculated based on the number of stillbirth and the estimated number of live birth for the country period from the UN World Population Prospects (\cite{wpp2019}). For observations with unknown $s_i$, we impute the maximum error of that source type.

\subsection{SBR model details}\label{sec-app-eqs}
This section includes the full description of the BHTSRM model, repeating the equations from the main text and adding priors used.
We denote $y_i$ as an observed SBR for country $c[i]$, in year $t[i]$, $\Theta_{c,t} = \log(\Omega_{c,t})$ as the log-transformed true SBR $\Omega_{c,t}$ for that country-year,  $s_i^2$ refers to its variance, $\psi_{j[i]}$ and $\sigma_{j[i]}^2$ to its source type-specific bias and variance, respectively, and $\hat{\gamma}_{d}$ and $\hat{\varphi}_{d}^2$ as definition-specific adjustment and variance. We assume 
\begin{eqnarray*}
\log(y_i)|\Theta_{c[i],t[i]}, \psi_{j[i]}, \sigma_{j[i]}^2 &\sim& N(\Theta_{c[i],t[i]} + \psi_{j[i]} + \hat{\gamma}_{d[i]}, s_i^2 + \hat{\varphi}_{d[i]}^2 + \sigma_{j[i]}^2),
\end{eqnarray*}
with
\begin{eqnarray*}
\Theta_{c,t} &=& \varsigma_c + \sum_k X_{k,c,t} \beta_k + \delta_{c,t},
\end{eqnarray*}
where $\varsigma_c$ refers to the country-specific intercept, $\sum_k X_{k,c,t} \beta_k$ refers to the linear regression function and $\delta_{c,t}$ refers to a temporal smoothing process. Prior distributions on source-type specific bias $\psi_j$ and variance $\sigma_{j}^2$, with source index $j = 1,2,3,4$ referring to administrative, HMIS, population-based studies, and survey data, respectively, are as follows: 
\begin{eqnarray*}
\psi_{1,2,3} &=& 0,\\
\psi_{4} &\sim& N^{-}(0,5^2),\\
\sigma_{1,2,3,4} &\sim& N^+(0,1). 
\end{eqnarray*}

 Country-specific intercepts $\varsigma_c$ are estimated hierarchically, with 
\begin{eqnarray*}
\varsigma_c|\eta_{r[c]},\sigma_{\varsigma}^2 &\sim& N(\eta_{r[c]},\sigma_{\varsigma}^2),\\
\eta_r|\xi_w,\sigma_{\eta}^2 &\sim& N(\xi_w,\sigma_{\eta}^2),
\end{eqnarray*}
where $\eta_{r[c]}$ refers to the regional mean, $\sigma_{\varsigma}^2$ to the across-country variance within regions, and $\xi_w$ to the global mean and $\sigma_{\eta}^2$ to the across-region variance. Vague priors were used for the global mean and variances:
\begin{eqnarray*}
\xi_w &\sim& N(2.5,2^2),\\
\sigma_{\varsigma},\sigma_{\eta} &\sim& N^+(0,1).
\end{eqnarray*}

The smoothing process is defined as follows:
\begin{eqnarray*}
\delta_{c,t} &=& \sum_{h=1}^H k_h(t)\alpha_{h,c},
 \end{eqnarray*}
 where $k_h(t)$ refers to the $h$-th spline function evaluated at time $t$ and $\alpha_{h,c}$ to its regression coefficient for country $c$, with $\frac{1}{H} \sum_h \alpha_{h,c} = 0$ and 
 \begin{eqnarray*}
\Delta \alpha_{h,c} &=& \alpha_{h,c} - \alpha_{h-1,c} \sim N(0,\sigma_{\Delta}^2),\\
\sigma_\Delta &\sim& U(0,3).
\end{eqnarray*}

Regression coefficients are estimated using horseshoe priors: 
\begin{eqnarray*}
\beta_k|\lambda_k,\tau,\rho &\sim&  N(0,\tau^2\tilde{\lambda}_k^2),\\
\tilde{\lambda}_k^2 &=& \frac{\rho^2\lambda_k^2}{\rho^2+\tau^2\lambda_k^2},\\
\lambda_k &\sim& C^+(0,1),\\
\tau &\sim& C^+ (0, 1),\\
\rho^2 &\sim& \text{Inv-Gamma}(2,8),
\end{eqnarray*}
where $\tau$ and $\rho$ are global shrinkage parameters, and the $\lambda_k$s are local (coefficient-specific) parameters.

\subsection{Definitional adjustments}\label{sec-app-def}

We write the true log-transformed SBR for observation $i$ under definition $d$, $\Theta_{c[i],t[i]}^{(d)}$, as follows:
\begin{align*}
\Theta_{c[i],t[i]}^{(d)} &= \Theta_{c[i],t[i]} + \kappa_i^{(d)}, 
\end{align*}
where $\Theta_{c,t}$ refers to the log-transformed SBR, such that $\kappa_i^{(d)}$ refers to the log-ratio of SBRs $\Omega$: 
\begin{align*}
\kappa_i^{(d)} =\log \left( \frac{\Omega_{c[i],t[i]}^{(d)}}{\Omega_{c[i],t[i]}}\right).
\end{align*}

Given that the number of stillbirths are small relative to live births, we can approximate $\kappa^{(d)}$ by the ratio of stillbirths. Specifically:
\begin{align*}
\Omega_{c,t}^{(d)} &= \frac{\Upsilon_{c,t}^{( d)}}{B_{c,t} + \Upsilon_{c,t}^{(d)}},\label{eq-ratiotonumbe}
\end{align*}
where $\Upsilon_{c,t}^{(d)}$ refers to the ``true'' stillbirth count associated with the true SBR, and $B_{c,t}$ the number of live births, with $\Upsilon_{c,t}^{(d)} << B_{c,t}$ such that \begin{align*}
\kappa_i^{(d)} =\log \left( \frac{\Omega_{c[i],t[i]}^{(d)}}{\Omega_{c[i],t[i]}}\right)
\approx \log \left( \frac{\Upsilon_{c[i],t[i]}^{( d)}}{\Upsilon_{c[i],t[i]})}\right).
\end{align*}

We obtain median estimate ${\gamma}_{d}$ and variance ${\varphi}_{d}^2$ of the predictive distribution for $\kappa_i^{(d)}$s for each alternative definition $d$ based on this approximation. Specifically, we set 
\begin{eqnarray}
\kappa_i^{(d)} &=& \log \left( \frac{\Upsilon_{c[i],t[i]}^{( d)}}{\Upsilon_{c[i],t[i]})}\right),
\label{eq-kappa}
\end{eqnarray}
and estimate its predictive distribution for each alternative definition.

\subsubsection*{Definitions containing the $\geq 28$ weeks definition}
For observations with definitions containing the $\geq 28$ weeks definition,  we denote $z_i$ as the number of stillbirths under the $\geq 28$ weeks definition, $z_i^{(d)}$ as the number of stillbirths under the alternative definition, and $i = 1, \dots, n_d$, where $n_d$ is the number of pairs available for definition $d$. We asume
\begin{eqnarray*}
z_i | \omega_i^{(d)} &\sim& \text{Binomial}(z_i^{(d)}, \omega_i^{(d)}),\\
\text{logit}(\omega_i^{(d)})|\mu_{\omega, d}, \sigma_{\omega, d}^2 &\sim& N(\mu_{\omega, d}, \sigma_{\omega, d}^2),\\
\end{eqnarray*}
where $\omega^{(d)}$ is the definition-specific probability of a stillbirth with gestational age of $\geq$ 22 or $\geq$ 24 weeks being born dead after 28 weeks, $\mu_{\omega, d}$ is the mean of the logit-transformed probabilities, and $\sigma_{\omega, d}$ the standard deviation. As per Equation~(\ref{eq-kappa}), we set $\kappa_i^{(d)} = \log(\Upsilon_{c[i],t[i]}^{( d)}/ \Upsilon_{c[i],t[i]})$, here
\begin{align*}
\kappa_i^{(d)} = \log(\Upsilon_{c[i],t[i]}^{( d)}/ \Upsilon_{c[i],t[i]}) = -\log(\omega_i^{(d)}).  
\end{align*}

We use vague prior for the mean and variance parameters:
\begin{eqnarray*}
\sigma_{\omega, d} &\sim& N^{+}(0,1),\\
\text{expit}(\mu_{\omega, d}) &\sim& U(0,1).
\end{eqnarray*}

\subsubsection*{Definitions overlapping with the $\geq 28$ weeks definition}
We denote $a_i$ as the count of stillbirths that satisfy the 28-week and alternative definition, $b_i$ as the counts of stillbirth with alternative definition rather than 28 weeks definition, and $c_i$ as the count of stillbirth with 28 weeks definition rather than alternative definition. We assume 
\begin{eqnarray*}
  (a_i,b_i,c_i)|(\omega_{i,a}^{(d)},\omega_{i,b}^{(d)},\omega_{i,c}^{(d)}) &\sim& Multinom\left(N_i,(\omega_{i,a}^{(d)},\omega_{i,b}^{(d)},\omega_{i,c}^{(d)})\right),
\end{eqnarray*}
where $\omega_{i,a}^{(d)}, \omega_{i,b}^{(d)}$ and $\omega_{i,c}^{(d)}$ refer to the probability of a stillbirth of the 28-week or alternative definition satisfying both definitions, the 28-week only, and the alternative definition only. 

We define $\Gamma_i^{(d)}$ to refer to the log-ratio of the definition-specific probabilities:
\begin{eqnarray*}
\Gamma_i^{(d)}  &=& \log\left(\frac{\omega_{i,a}^{(d)}+\omega_{i,b}^{(d)}}{\omega_{i,a}^{(d)}+\omega_{i,c}^{(d)}}\right).
\end{eqnarray*}
With Equation~(\ref{eq-kappa}), we find that $\kappa$ equals the log-ratio of the definition-specific probabilities $\Gamma_i^{(d)}$:
\begin{align*}
\kappa_i^{(d)} = \log(\Upsilon_{c[i],t[i]}^{( d)}/ \Upsilon_{c[i],t[i]}) = \log\left(\frac{\omega_{i,a}^{(d)}+\omega_{i,c}^{(d)}}{\omega_{i,a}^{(d)}+\omega_{i,b}^{(d)}}\right) = \Gamma_i^{(d)}.
\end{align*}

We assume that the $\Gamma_i^{(d)}$s are normally distributed,
\begin{eqnarray*}
\Gamma_i^{(d)}|\mu_{\Gamma,d },\sigma_{\Gamma, d}^2 &\sim& N(\mu_{\Gamma,d },\sigma_{\Gamma, d}^2),\label{eq-Gamma}
\end{eqnarray*}
with $\mu_{\Gamma,d }$ and $\sigma_{\Gamma, d}^2$ referring to the across-setting mean and variance of the log-ratios. To guarantee that the estimation results in sets of $\omega_{i,a}^{(d)}, \omega_{i,b}^{(d)}$ and $\omega_{i,c}^{(d)}$ that add up to one, we introduce the constraint $\frac{1}{1+\exp(\Gamma_i^{(d)})} < \omega_{i,b}^{(d)}+\omega_{i,c}^{(d)}< \frac{1}{\max\{1, \exp(\Gamma_i^{(d)})\}}$. We incorporate this constraint through a prior on the sum: 
\begin{eqnarray*}
\left(\omega_{i,b}^{(d)}+\omega_{i,c}^{(d)}\right)|\Gamma_i^{(d)} &\sim& U\left(\frac{1}{1+\exp(\Gamma_i^{(d)})}, \frac{1}{\max\{1, \exp(\Gamma_i^{(d)})\}} \right).
\end{eqnarray*}
We use vague priors for the mean and variance parameters:
\begin{eqnarray*}
\sigma_{\Gamma, d} &\sim& N^{+}(0,1),\\
\mu_{\Gamma,d } &\sim& N(0, 20).
\end{eqnarray*}

\subsection{Model comparison and sensitivity analyses}\label{sec-modelcompare}
\begin{figure}[htbp]
    \centering
       \includegraphics[width=1\textwidth]{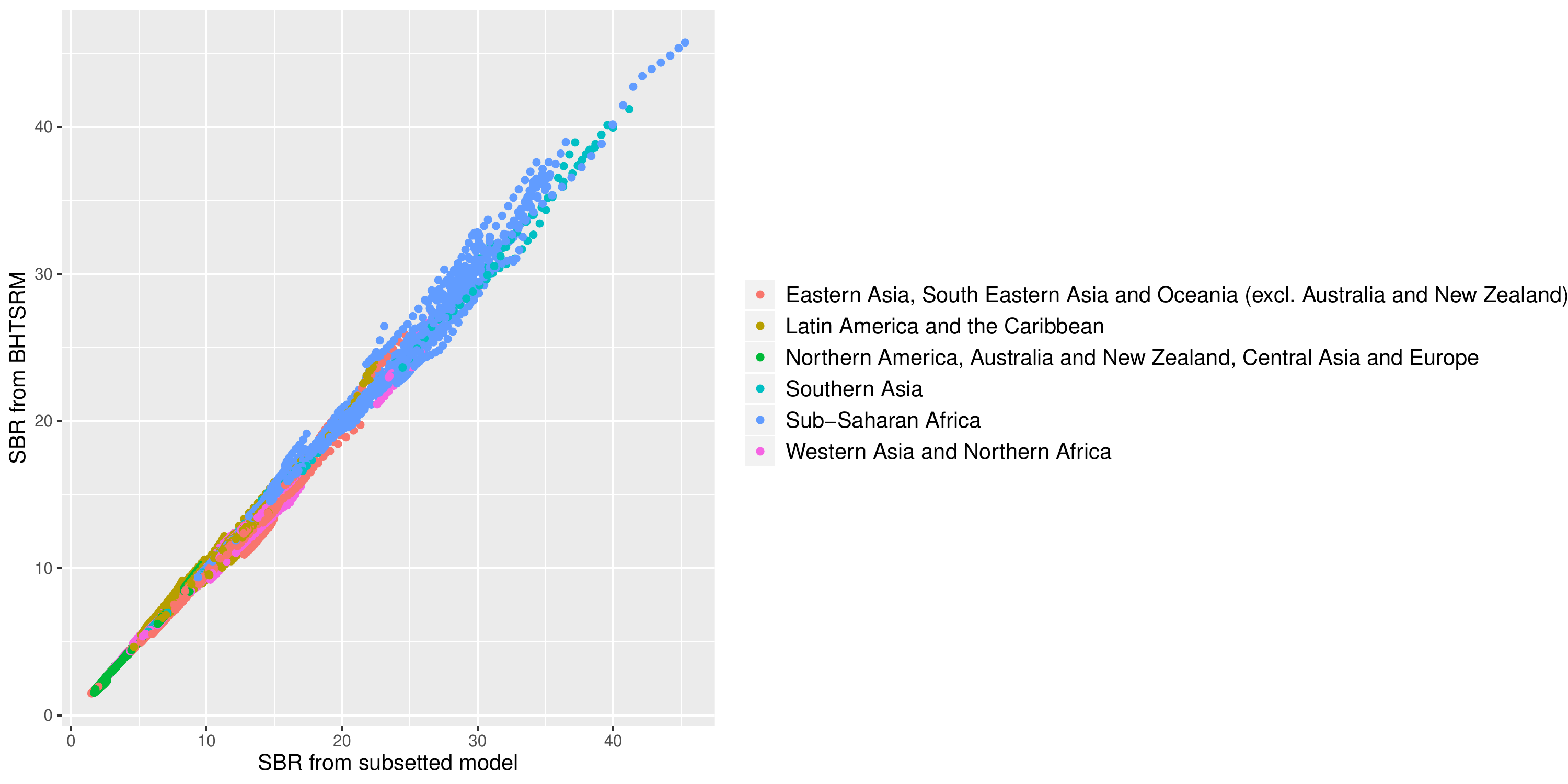}\\
       \includegraphics[width=1\textwidth]{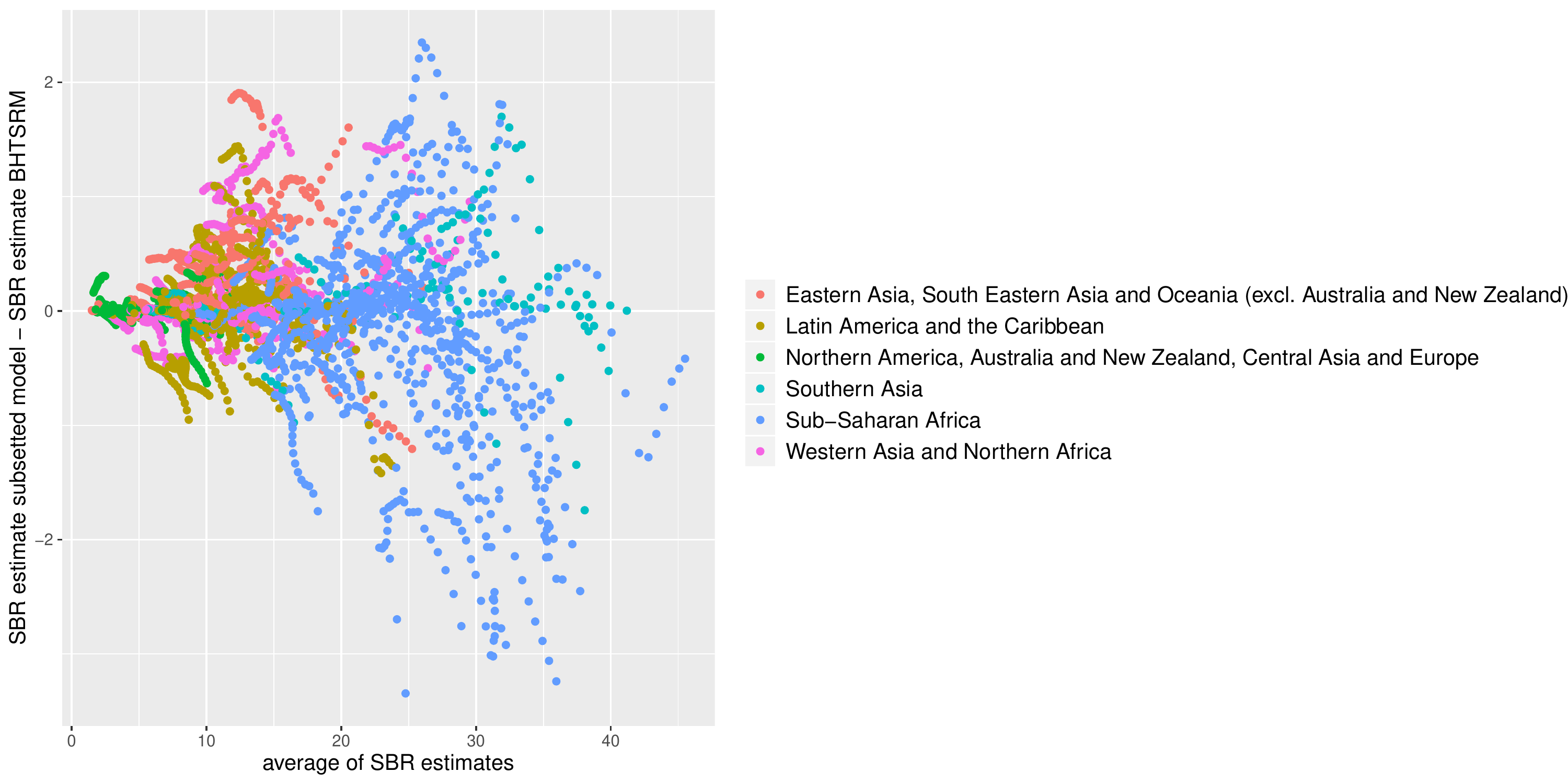}\\
       \caption{Comparison between SBR estimates from the BHTSRM and subsetted model. Top: estimated SBR from BHTSRM against SBR from subsetted model. Bottom: difference in SBR between subsetted model and BHTSRM against the average of the two estimates.}
       \label{fig-est_comp}
\end{figure}

We compare model fits based on the model using horseshoe priors and subsetted model fit. The comparison of the estimates of the regression coefficients are presented in Table~\ref{tab-com_cov}. 
Figure~\ref{fig-est_comp} provides country fits of BHTSRM and subsetted model. Estimates are not substantially different. 

\begin{table}[ht]
\centering
\begin{tabular}{l|cc|cc}
  \hline
  \multicolumn{1}{c}{} & \multicolumn{2}{c}{BHTSRM} & \multicolumn{2}{c}{Subsetted model} \\  
  covariates & estimate & sd & estimate & sd \\ 
  \hline
log(NMR) & 0.414 & 0.040 & 0.401 & 0.037  \\ 
  log(GNI) & -0.102 & 0.056 & -0.116 & 0.038 \\ 
  log(lbw) & 0.078 & 0.033 & 0.097 & 0.028  \\ 
  edu & -0.037 & 0.031 & -0.062 & 0.029  \\ 
  csec & -0.027 & 0.025 & -0.047 & 0.024   \\ 
  anc4 & -0.025 & 0.029 & -0.052 & 0.030  \\ 
pab      & -0.018 & 0.015     & &  \\ 
abr      & -0.017 & 0.035     &  &  \\ 
urban    & -0.012 & 0.029     &  & \\ 
gini     & 0.010  & 0.020     &  & \\ 
sab      & -0.010 & 0.027     &  &  \\ 
anc1     & -0.009 & 0.022     &  &  \\ 
mmr      & 0.003  & 0.038     &  &  \\ 
pfpr     & -0.002 & 0.017     &  &  \\ 
gdp      & 0.001  & 0.025     & &  \\ 
gfr      & 0.000  & 0.025     &  &  \\ 
  \hline
  \end{tabular}\caption{Overview of estimates for regression coefficients under different model fits (posterior mean and posterior standard deviation). Details on covariates is given in Appendix Table~\ref{tab-cov_inf}.}
\label{tab-com_cov}
\end{table}

For comparing expected log pointwise predictive density (ELPD), we add one additional model to the comparison, based on an alternative choice of hyperparameters for the horseshoe prior based on \cite{piironen_sparsity_2017}. 
For standard regression models with $y_i \sim N((\bm{X}\bm{\beta})_i, \sigma^2)$, Piironen and Vehtari (2017) propose to set the scale parameter $\tau_0$ in the prior for $\tau$ as follows:
\begin{eqnarray*}
\tau \sim C^{+}(0,\tau_0),\\
\tau_0 = \frac{p_0}{D-p_0}\frac{\sigma}{\sqrt{n}},
\end{eqnarray*}
where $p_0$ is the guess of number of relevant predictors, $D$ is the total number of predictors, $\sigma$ is the standard deviation of observation $\log(y)$, and $n$ is the number of observations. We do not follow this recommendation because our modeling context differs from the one where this setting was explored, i.e. our setting includes heteroskedasticity of observations and the regression model is combined with a temporal smoothing term. We obtain a model fit based on the recommendation as a sensitivity test. Specifically, we obtain the fit for $p_0 = 5$, $D= 16$, $\sigma = 0.094$ (the median standard deviation across observations), and $n= 1531$, corresponding to $\tau_0 = 0.001$.

\begin{figure}[htbp]
    \centering
       \includegraphics[page=3,width=0.85\textwidth,height=0.3\textheight]{fig/psis.pdf}\\
       \includegraphics[page=2,width=0.85\textwidth,height=0.3\textheight]{fig/psis.pdf}\\
       \includegraphics[page=1,width=0.85\textwidth,height=0.3\textheight]{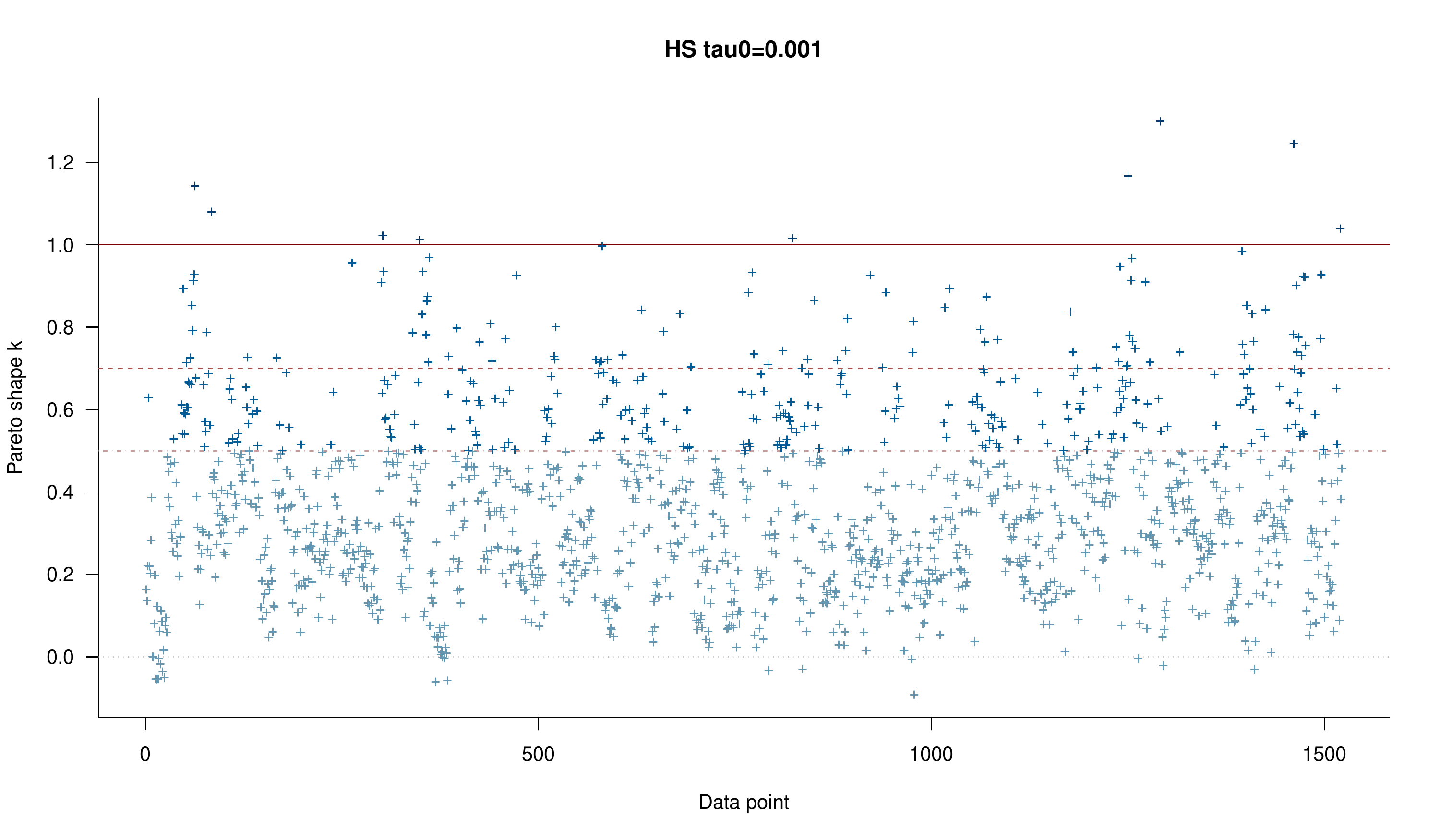}\\
    \caption{PSIS diagnostic plots for three models as specified in the Appendix Section~\ref{sec-modelcompare}. }
    \label{fig:psis}
\end{figure}

Table \ref{tab-waic} summarizes the differences in ELPD among the three models. 
There are no significant differences among the ELPDs. The PSIS diagnostics (see Figure \ref{fig:psis}) indicate that there are some data points with large Pareto $k$ values. The percentage of points with $k > 0.7$ (suggesting outlying and possibly influential points) is smallest for the model based on horseshoe prior with $\tau_0=1$ (6.6\%).

\begin{table}[htbp]
\centering
\begin{tabular}{l|cc|cc}
\multirow{2}{*}{Models}   & \multicolumn{2}{c|}{95\% CI for difference in ELPD}    & \multicolumn{2}{c}{elpd} \\ \cline{2-5} 
                          & HS $\tau_0=0.001$ & subsetted model & elpd\_loo     & SE       \\ \hline
HS $\tau_0=1$                & (-80.7,94.9)        & (-78.8,95.2)              & 1084.4        & 45.5     \\
HS $\tau_0=0.001$            &                      & (-4.3,6.5)                & 1091.5        & 45.3     \\
subsetted model &                &                                            & 1092.6        & 44.9     \\ \hline
\end{tabular}
\caption{Model comparison based on expected log pointwise predictive density. }\label{tab-waic}
\end{table}

\clearpage

\subsection{Covariates}

Covariates are described in Table \ref{tab-cov_inf}.

\begin{longtable}{|p{0.05\textwidth}|p{0.3\textwidth}|p{0.35\textwidth}|p{0.35\textwidth}|}
\caption{Candidate covariates with its sources and methodology} \label{tab-cov_inf} \\

\hline \multicolumn{1}{|c|}{\textbf{Var}} & \multicolumn{1}{c|}{\textbf{Source}} & \multicolumn{1}{c|}{\textbf{Definition}} & \multicolumn{1}{c|}{\textbf{Methodology Notes}} \\ \hline 
\endfirsthead

\multicolumn{3}{c}%
{{\bfseries \tablename\ \thetable{} -- continued from previous page}} \\
\hline \multicolumn{1}{|c|}{\textbf{Var}} & \multicolumn{1}{c|}{\textbf{Source}} & \multicolumn{1}{c|}{\textbf{Definition}} & \multicolumn{1}{c|}{\textbf{Methodology Notes}} \\ \hline 
\endhead

\hline \multicolumn{3}{|r|}{{Continued on next page}} \\ \hline
\endfoot

\hline \hline
\endlastfoot
	abr & United Nations Department of Economic and Social Affairs (DESA), Population Division
United Nations Population Fund (UNFPA). Data are based on DHS, MICS and other national household surveys & Adolescent Birth Rate (number of live births to adolescent women per 1,000 adolescent women) & Extrapolated to 2019 assuming a flat trend, linear interpolation applied when data between 2000-2019 unavailable, imputation using regional year data for countries without any available data, and smoothing applied. \\ \hline
	anc1 & UNICEF/WHO. Data are based on DHS, MICS and other national household surveys & Antenatal care 1+ visit - Percentage of women (age 15–49) attended at least once during pregnancy by skilled health personnel. & Extrapolated to 2019 assuming a flat trend,  linear interpolation applied when data between 2000-2019 unavailable, imputation using regional year data for countries without any available data, and smoothing applied. \\ \hline
	anc4 & UNICEF/WHO. Data are based on DHS, MICS and other national household surveys & Antenatal care 4+ visits - Percentage of women (age 15–49)  attended at least four times during pregnancy by any provider.  & Extrapolated to 2019 assuming a flat trend,  linear interpolation applied when data between 2000-2019 unavailable, imputation using regional year data for countries without any available data, and smoothing applied. \\ \hline
	csec & UNICEF. Data are based on DHS, MICS and other national household surveys & C-section rate - Percentage of deliveries by Caesarian section.  & Extrapolated to 2019 assuming a flat trend,  linear interpolation applied when data between 2000-2019 unavailable, imputation using regional year data for countries without any available data, and smoothing applied. \\ \hline
	gfr & United Nations Department of Economic and Social Affairs (DESA), Population Division, World Population Prospects 2019 Edition & General fertility rate. Number of live births divided by the female population age 15-49 years. & No additional processing applied. \\ \hline
	gdp & World Bank & World Bank Gross domestic product per capita & No additional processing applied. \\ \hline
	gini & World Bank, Development Research Group. Data are based on primary household survey data obtained from government statistical agencies and World Bank country departments. & Gini index measures the extent to which the distribution of income (or, in some cases, consumption expenditure) among individuals or households within an economy deviates from a perfectly equal distribution. & Extrapolated to 2019 assuming a flat trend,  linear interpolation applied when data between 2000-2019 unavailable, imputation using regional year data  for countries without any available data, and smoothing applied. \\ \hline
	gni & World Bank, International Comparison Program &  Gross national income per capita & Extrapolated to 2019 assuming a flat trend, linear interpolation applied when data between 2000-2019 unavailable, and imputation using regional year data for countries without any available data. \\ \hline

	lbw & UNICEF/WHO  estimates, 2019 Edition. Data based on vital registration data and national household surveys & Percentage of live births that weighted less than 2500 grams. & Extrapolated to 2019 assuming a flat trend, linear interpolation applied when data between 2000-2019 unavailable, imputation using regional year data for countries without any available data, and smoothing applied. \\ \hline
	edu & United Nations Development Programme. Data are based Barro and Lee (2013), UNESCO Institute for Statistics (2013).

 & Average number of years of education received by people ages 25 and older, converted from educational attainment levels using official duration of each level. & Extrapolated to 2019 assuming a flat trend, linear interpolation applied when data between 2000-2019 unavailable, imputation using regional year data for countries without any available data, and smoothing applied. \\ \hline
	mmr & UN MMEIG estimates, 2019 edition. Data are based on DHS, MICS and other national household surveys & The number of maternal deaths during a given time period per 100,000 live births during the same time period.  & Extrapolated to 2019 assuming a flat trend, and imputation using regional year data  for countries without any available data. \\ \hline
	nmr & UN IGME, 2019 Edition. Data are based on from vital registration, household survey and population census. & Probability of dying in the first 28 days of life, expressed per 1,000 live births. & No additional processing applied. \\ \hline
	pab & UNICEF/WHO. Data based on administrative reporting and TT coverage surveys. & Percentage of pregnant women protected by tetanus toxoid containing vaccines (TTCV) who would give birth to a child protected against tetanus as a result of maternal transfer of antibodies through the placenta. & Extrapolated to 2019 assuming a flat trend, linear interpolation applied when data between 2000-2019 unavailable, imputation using regional year data for countries without any available data, and smoothing applied. \\ \hline
	pfpr & Malaria Atlas Project, estimates, 2019 edition. Data based on national household surveys, routine surveillance systems, and geographic and climate data & Plasmodium falciparum parasite rate. & Extrapolated to 2019 assuming a flat trend, linear interpolation applied when data between 2000-2019 unavailable,  and smoothing applied. \\ \hline
	sab & UNICEF/WHO estimates, 2019 edition. Data based on admin records, DHS, MICS and other national household surveys & The proportion of births attended by skilled health personnel.& Extrapolated to 2019 assuming a flat trend,  linear interpolation applied when data between 2000-2019 unavailable, and imputation using regional year data  for countries without any available data. \\ \hline
	urban & United Nations Department of Economic and Social Affairs, Population Division, World Urbanization Prospects 2018 & Percentage of population living in urban areas. & No additional processing applied.

\end{longtable}

\clearpage 



\end{document}